\documentclass[a4paper,fleqn,usenatbib]{mnras}

\usepackage{newtxtext,newtxmath}

\usepackage[T1]{fontenc}
\usepackage{ae,aecompl}

\usepackage{graphicx}	
\usepackage{amsmath}	
\usepackage{amssymb}	
\usepackage{pdfpages}

\title[Sulphur chemistry in dark clouds]{On the reservoir of sulphur in dark clouds : chemistry and elemental abundance reconciled}

\author[T. Vidal et al.]{
Thomas H. G. Vidal,$^{1}$\thanks{E-mail: thomas.vidal@u-bordeaux.fr}
Jean-Christophe Loison$^{2,3}$,
Adam Yassin Jaziri$^{1}$,
Maxime Ruaud$^{1}$,\newauthor
Pierre Gratier$^{1}$
and Valentine Wakelam$^{1}$
\\
$^{1}$Laboratoire d'astrophysique de Bordeaux, Univ. Bordeaux, CNRS, B18N, allée Geoffroy Saint-Hilaire, 33615 Pessac, France\\
$^{2}$Univ. Bordeaux, ISM, UMR 5255, F-33400, Talence, France\\
$^{3}$CNRS, ISM, UMR 5255, F-33400, Talence, France\\
}

\date{Accepted XXX. Received YYY; in original form ZZZ}

\pubyear{2016}

\begin{document}
\label{firstpage}
\pagerange{\pageref{firstpage}--\pageref{lastpage}}
\maketitle

\begin{abstract}
Sulphur-bearing species are often used to probe the physical structure of star forming regions of the interstellar medium, but the chemistry of sulphur in these regions is still poorly understood. In dark clouds, sulphur is supposed to be depleted under a form which is still unknown despite numerous observations and chemical modeling studies that have been performed. In order to improve the modeling of sulphur chemistry, we propose an enhancement of the sulphur chemical network using experimental and theoretical literature. We test the effect of the updated network on the outputs of a three phases gas-grain chemical model for dark cloud conditions using different elemental sulphur abundances. More particularly, we focus our study on the main sulphur reservoirs as well as on the agreement between model predictions and the abundances observed in the dark cloud TMC-1 (CP). Our results show that depending on the age of the observed cloud, the reservoir of sulphur could either be atomic sulphur in the gas-phase or HS and H$_2$S in icy grain bulks. We also report the first chemical model able to reproduce the abundances of observed S-bearing species in TMC-1 (CP) using as elemental abundance of sulphur its cosmic value.
\end{abstract}




\section{Introduction}

Since the first detection of a S-bearing molecule in the interstellar medium \citep[CS,][]{Penzias71}, sulphur has become a subject of interest for many astrochemists. S-bearing molecules are indeed often used to probe the physical structure of star-forming regions \citep{Lada91,Plume97,Viti01,Sakai14,Podio15} and have been proposed as chemical clocks for hot cores \citep{Charnley97,Hatchell98,Wakelam04}. But sulphur chemistry in the dense interstellar medium has also been at the centre of a puzzling issue for many years, namely the sulphur depletion problem \citep{Ruffle99}. Unlike most of the other elements, in the diffuse medium, the gas-phase abundance of atomic sulphur is observed to be constant with cloud density, around its cosmic value of a few $10^{-5}$ \citep[see for instance][]{Jenkins09}. However, in dark clouds the total abundance of detected S-bearing molecules only accounts for 0.1\% of the cosmic abundance of atomic sulphur \citep{Tieftrunk94,Charnley97}. Therefore, the main reservoirs of sulphur in dark clouds are still unknown. One could argue that sulphur is in its atomic form, i.e. non observable. However, assuming an elemental abundance of sulphur in chemical models as high as its cosmic value produces predicted abundances of observable S-bearing molecules much higher than the observed one. Consequently, modelers usually assume that the elements (sulphur but other heavy elements as well) are depleted compared to their cosmic reference values and they use the so-called 'depleted' abundance of sulphur of a few $10^{-8}$ \citep{Wakelam08}. \\

The main hypothesis to explain this missing sulphur is that it is depleted onto interstellar grains. In cometary ices, which are thought to present chemical similarities with the ices processed during hot core formation \citep{Irvine00}, H$_2$S is the most abundant S-bearing molecule, at the level of 1.5\% compared to water \citep{Morvan00}. More recently, \cite{Holdship16} studied the properties of H$_2$S in the low-mass protostar L1157-B1 and found that a significant fraction of the sulphur is likely to be locked into the form of H$_2$S on the grains. Chemical models also predict that, in the dense interstellar medium, atomic sulphur would stick on grains and be mostly hydrogenated to form H$_2$S. To this day however only OCS \citep{Palumbo97} and SO$_2$ \citep{Boogert97} have been likely identified in icy grain bulks in dense molecular clouds surrounding high-mass protostars and their estimated total abundance does not account for the missing sulphur. Upper limits for the column density of H$_2$S in icy grain bulks have been derived by \citet{Smith91}, notably towards the line of sight of three late-type field stars lying behind the Taurus dark cloud, but these are also too low for H$_2$S to be the reservoir of sulphur in dark clouds.

Laboratory experiments coupled with chemical models have recently brought new insight into the problem by studying the irradiation of H$_2$S interstellar ice analogs by energetic protons and UV photons  by \citet{Garozzo10} and \citet{Jimenez11}. Both studies found that solid H$_2$S was easily destroyed to form other species such as OCS, SO$_2$, CS$_2$ and H$_2$S$_2$. Subsequently, a corresponding network has was proposed by \citet{Druard12} to reproduce these experiments but the authors showed that sulphur was converted into other forms, mostly H$_2$S$_2$ and H$_2$S$_3$, at higher temperatures ($>20$ K) than those found in dark clouds (around 10 K). \\

In this article, we propose an enhancement of the sulphur chemical network for dark clouds simulations motivated by the recent observations of HNCS and HSCN in TMC-1 (CP) \citep{Adande10} and of CH$_3$SH in IRAS 16293-2422 \citep{Majumdar16}. Our network also includes the network proposed by \citet{Druard12}. We test the effect of the updated network on the outputs of a gas-grain chemical model for dark cloud conditions using different sulphur elemental abundances. More particularly, we focus our study on the main sulphur reservoirs as well as on the agreement between model predictions and the abundances observed in the dark cloud TMC-1 (CP). The gas-grain model we used as well as the modifications of the chemical network are presented in sections 2 and 3. Sections 4 and 5 describe our new modeling results and include comparisons with the previous version of the network as well as comparisons with observations in the dark cloud TMC-1 (CP). We discuss and conclude on our results in the last section.

\section{Model}

The model we use to assess the impact of our new network on the predicted S-bearing species abundances of dark clouds, is the latest version of the Nautilus chemical model, described in \citet{Ruaud16}. This model allows us to compute the evolution of chemical abundances for a given set of physical and chemical parameters. Recently updated, it simulates a three phases chemistry including gas-phase, grain surface and grain bulk chemistries, along with the possible exchanges between the phases. These exchanges are:  the adsorption of gas-phase species onto grain surfaces, the thermal and non-thermal desorption of species from the grain surface into the gas-phase, and the surface-bulk and bulk-surface exchange of species. Our reference model for this article uses the formulation of chemical desorption processes depicted in \citet{Garrod07}. They consider that for each surface reactions leading to a single product, a part of the energy released by the reaction can contribute to the desorption of the product in the gas-phase using the  Rice-Ramsperger-Kessel theory. The fraction of the product desorbed in the gas-phase depends on the binding energy of the product to the surface, the enthalphy of the reaction, and the fraction of the released energy that is lost to the surface. In our case, we use a $a$ parameter of 0.001, which produces approximately a 1\% efficiency evaporation for all species. Moreover, the grain chemistry takes into account the standard direct photodissociation by UV photons along with the photodissociation induced by secondary UV photons introduced by \citet{Prasad83}. These processes are effective on the surface as well as in the bulk of the grains. The model also takes into account the newly implemented competition between reaction, diffusion and evaporation as suggested by \citet{Chang07} and \citet{Garrod11}. The diffusion energies of each species are computed as a fraction of their binding energies. We take for the surface a value of this ratio of 0.4 as suggested by experiments and theoretical work made on H \citep[see][and reference therein]{Ruaud16}, CO and CO$_2$ \citep[see][]{Karssemeijer14}. This value is then extrapolated to every species on the surface. For the bulk we take a value of 0.8 \citep[see also][]{Ruaud16}. The reference (gas-phase and grains) network is \textit{kida.uva.2014} \citep[see][]{Wakelam15}, the same one as in \citet{Ruaud16}.\\
The set of physical parameters used throughout this article is the commonly used dark cloud parameter configuration, namely a gas and dust temperature of 10 K, an atomic hydrogen total density of $2\times10^4$ cm$^{-3}$, a cosmic ionization rate of $1.3\times10^{-17}$~s$^{-1}$, and a visual extinction of 15 mag. All abundances are expressed with respect to the total H density. Our set of initial abundances is summarized in Table \ref{tab_1}. We start with all species in their atomic (or ionized) form, except for hydrogen which is assumed to be entirely in its molecular form.

\begin{table}
\caption{Initial abundances. *$a(b)$ stands for $a\times10^b$}
	\begin{center}
		\begin{tabular}{l r r}
		\hline
		\hline
   		Element & $n_i/n_H$* & References \\
   		\hline
		H$_2$    & 0.5             &    \\
   		He          & 0.09           & 1 \\
		N            & 6.2(-5)       &  2 \\
		O            & 2.4(-4)       &  3 \\
		C$^+$    & 1.7(-4)       &  2 \\
		S$^+$    & 8.0(-8)       &  4 \\
		Si$^+$   & 8.0(-9)       &  4 \\
		Fe$^+$  & 3.0(-9)       & 4 \\
		Na$^+$  & 2.0(-9)       & 4 \\
		Mg$^+$ & 7.0(-9)       & 4 \\
		P$^+$   & 2.0(-10)     & 4 \\
		Cl$^+$  & 1.0(-9)       & 4 \\
		F           & 6.7(-9)     & 5 \\
		\hline
 		\end{tabular}
	\end{center}
	\medskip{(1) \citet{Wakelam08}, (2) \citet{Jenkins09}, (3) \citet{Hincelin11}, (4) Low-metal abundances from \citet{Graedel82}, (5) Depleted value from \citet{Neufeld05}}
  	\label{tab_1}
\end{table}

\section{Modification of the sulphur network}

To update the sulphur chemistry network, we first examine the existing KIDA network \citep[\textit{kida.uva.2014},][]{Wakelam15} looking systematically at the possible reactions between S and S$^+$ with the most abundant species in dense molecular clouds (CO, CH$_4$, C$_2$H$_2$, c-C$_3$H$_2$) as well as the potential reactions between sulphur compounds and the most abundant reactive species in molecular clouds (C, C$^+$, H, N, O, OH, CN). We found that various neutral reactions were missing from KIDA (or other databases) such as C + H$_2$S, C + H$_2$CS and S + l,c-C$_3$H. When previous experimental or theoretical studies exist, we use them to update the KIDA database when necessary. However, there are relatively few kinetics data for the reactions of sulphur compounds. To estimate the unknown rate constants to be used in the network, we use a methodology developed in appendix A (see supplementary material online) including new DFT and ab-initio calculations on the reactions H + CS, H + H$_2$CS, H + C$_2$S, H + HNCS, H + HSCN, NH + CS, NH$_2$ + CS, C$_2$H$_3$ + CS, O + C$_3$S, S + c-C$_3$H$_3^+$, S + c-C$_3$H$_2$, N + HCS. All calculations have been performed for gas phase reactions. The hydrogenation reaction barriers are used in the surface chemistry without changes. The rate constant of the NH + CS, NH$_2$ + CS, C$_2$H$_3$ + CS, O + C$_3$S reactions are presented in appendix B (see supplementary material online). It should be noted that there are often large uncertainties on rate constant values and branching ratios for the sulphur chemical network. This is particularly true for the branching ratios of dissociative recombination (DR). Most of them are deduced from general rules, developed in appendix A, by analogy with similar oxygenated compounds, but DR of HCS$^+$ leads mainly to S + CH \citep{Montaigne05} while DR of HCO$^+$ leads mainly to H + CO \citep{Hamberg14}. This may be critical for H$_2$CS (and also HC$_3$S) production if DR of  CH$_3$SH$^+$ leads mainly to C-S bond breaking in contrast to DR of CH$_2$OH$^+$ which preserves the C-O bond \citep{Hamberg07}.\\

In this work we introduce three isomers for HNCS (HNCS, HSCN, HCNS) and five for the protonated forms. We use the work of \citet{Gronowski14} to describe the ionic chemistry but, unlike them, we do not consider the CSH$^+$ isomer but only the HCS$^+$ one. Indeed, the CSH$^+$ $\to$ HCS$^+$ isomerization barrier is only 5.4 kJ/mol \citep{Puzzarini05} and as the reaction producing CSH$^+$ (CS + H$_3$$^+$) is exothermic by 78 kJ/mol, most of the CSH$^+$ should isomerize into HCS$^+$. The comparison with HOC$^+$ is not relevant as the COH$^+$/HCO$^+$ isomerization barrier is equal to 150 kJ/mol \citep{Nobes81}. We carefully searched for other reactions producing HNCS isomers, performing various theoretical calculations on the H-H-C-N-S system. For the NH$_2$ + CS reaction, we found a small barrier (+3.0 kJ/mol) at the CCSD(T)/cc-pVQZ//MP2/cc-pVTZ level (this barrier being submerged by -2.4 kJ/mol at the M06-2X/cc-pVTZ level but this method is thought to be less precise and usually slightly underestimates barrier values). We calculate the rate constant for this reaction using the CCSD(T) values leading to a negligible value at low temperature. We also include various reactions on the grain surface. Indeed, it appears that the s-N + s-HCS reaction is, alongside the following reaction mechanism:

\begin{eqnarray}
	\text{HCS}^+ + \text{NH}_2 \to \text{H} + \text{H$_2$NCS}^+ \xrightarrow{\text{e-}}  \text{HNCS/HSCN} + \text{H} 
\end{eqnarray}

the main source of HNCS isomers. The first step of this reaction is very likely N-C bond formation without a barrier through the pairing up of free electrons of the N atom and the HCS radical leading to HC(N)S formation. This species is indeed the transition state linking HNCS and HSCN. The branching ratio between HNCS and HSCN formation can be deduced from statistical theory. HNCS/HSCN production is estimated to be proportional to the density of vibrational states of each isomer near the effective barrier to isomerization in a similar manner to HCN/HNC formation in DR of HCNH$^+$ \citep{Herbst00}. Using the MESMER program \citep{Glowacki12} for the calculations of the ro-vibrational density of states near the isomerization barrier (see appendix A).\\

For the loss reactions we consider no barrier for the reactions of HNCS isomers with carbon atoms taking into account the high reactivity of carbon atoms with unsaturated closed shell molecules \citep[see Table 1 in][]{Loison14}. We neglect the reactions of HNCS isomers with H, N and O atoms in the gas phase either from M06-2X/cc-pVTZ calculations (see appendix A) or by comparison with similar reactions with HNCO isomers. For He$^+$ and H$^+$ reactions with HNCS isomers we consider, by comparison with similar reactions \citep{Anicich03}, a charge transfer mechanism as the main process leading to highly excited species followed by dissociation. For C$^+$ reactions with HNCS isomers, the charge exchange is usually a minor channel \citep{Anicich03}. We consider then that these reactions lead mainly to the most exothermic exit channel involving the fewest steps.  For protonation reactions we use the work of \citet{Gronowski14} leading to five H$_2$NCS$^+$ isomers. The DR rate constant of the five H$_2$NCS$^+$ isomers is assumed to be equal to the ion-electron collision rate estimated to be equal to $3\times10^{-7}\times(T/300)^{-0.5}$ considering the size of the cations \citep{Fournier13,Florescu06}. For the branching ratio of the DR of H$_2$NCS$^+$ and HNCSH$^+$, a critical parameter for the HNCS/HSCN ratio, we follow the same procedure as used for the DR of HCNH$^+$ \citep{Herbst00} using the MESMER program for the calculations of the ro-vibrational density of states near the isomerization barrier (see appendix A). Then, the DR reactions of H$_2$NCS$^+$ and HNCSH$^+$ leads to similar amounts of HNCS and HSCN. \\

Overall, we added (or reviewed in the case of reactions) 46 S-bearing species to the network along with 478 reactions in the gas-phase, 305 reactions on the grain surface and 147 reactions in the grain bulk (see the table in appendix B). The newly introduced species are listed in Table \ref{tab_1b}. Among these species, H$_2$S$_3$ and S$_n$ ($n$ = 3 to 8) are only present on the grains, i.e. they are formed on the surfaces and not allowed to evaporate \citep[see also][]{Druard12}. At 10K, these species are not efficiently formed so we do not want to add any gas-phase routes for these species as this chemistry is not known. It should be noted that our enhanced network includes the chemical schemes for carbon chains proposed in \citet{Wakelam15b}, \citet{Loison16}, \citet{Hickson16} and Loison et al. (in preparation).

\begin{table}
\caption{Table of the 46 S-bearing species added to the network.}
	\begin{center}
		\begin{tabular}{c c}
		\hline
		\hline
   		\multicolumn{2}{c} {Neutrals}\\
   		\hline
		CS$_2$         &  HNCS\\
		H$_2$S$_3$* &  HSCN\\
		S$_3$*            &  H$_2$C$_3$S\\
		S$_4$*           &  HNCHS\\
		S$_5$*	      & HSCHN\\
		S$_6$*           & HNCSH\\
		S$_7$*	      & NH$_2$CHS\\
		S$_8$*	      & H$_2$C$_2$S\\
		HC$_3$S       & NH$_2$CS\\
		HC$_2$S       & NH$_2$CH$_2$S\\
		CH$_3$S       & NH$_2$CH$_2$SH\\
		CH$_3$SH      & HSO\\
		CH$_2$SH    & HNCHSH\\
		HCNS	      & NH$_2$CHSH\\
		\hline
		\hline
		\multicolumn{2}{c} {Ions}\\
		\hline
		CS$_2^+$                    & HSCN$^+$ \\
		HCS$_2^+$                  & H$_2$CNS$^+$\\
		H$_2$C$_2$S$^+$      & H$_2$NCS$^+$\\
		H$_2$C$_3$S$^+$      & H$_2$SCN$^+$\\
		CH$_3$S$^+$    	    & HCNSH$^+$\\
		CH$_3$SH$^+$     	    & HNCSH$^+$\\
		CH$_3$SH$_2^+$	    & CH$_3$CS$^+$\\
		HCNS$^+$		    & NH$_2$CHSH$^+$\\
		HNCS$^+$		    & NH$_3$CH$_2$SH$^+$\\ 		    
		\hline
 		\end{tabular}
	\end{center}
	\medskip{* These species are not allowed to evaporate in the gas-phase (see text).}
  	\label{tab_1b}
\end{table}

\section{Impact of the new network on the chemical model of dark clouds} \label{chem_study}

In this section we highlight the impact of our enhanced sulphur network on the outputs of the Nautilus chemical code, configured for dark clouds physical conditions. We first present the abundance evolution of the most abundant sulphur-bearing species we obtain from the model in this configuration. Then we highlight our results on the newly implemented sulphur-bearing species. Finally, we compare the outputs obtained with our new network with those obtained with the nominal network for the same model configuration. 

\subsection{Abundances of the main sulphur-bearing species} \label{reservoirs}

In the following we study the abundances of the main sulphur-bearing species obtained with our enhanced network, both in the gas phase and on the grains. We choose these species since they contain more than 5 \% of the elemental sulphur at one point in our simulation.

\subsubsection{Main gas-phase species: S, CS and SO}

Figure \ref{fig_1} (a) shows the time evolution of the main sulphur-bearing species in the gas-phase: atomic sulphur (S), carbon monosulfide (CS) and sulphur monoxide (SO). In our simulation, sulphur is initially in the form of S$^+$ and between $7.7\times10^3$ and $4.6\times10^5$ years, S becomes the main sulphur reservoir (including species on grains). During this period, its abundance reaches a maximum at $10^5$ years when it contains up to 73\% of the elemental sulphur. The growth of the atomic S abundance up to this maximum seems to be mainly caused by the electronic recombination mechanisms of S$^+$, as well as of CS$^+$ and HCS$^+$. The last two are related to S$^+$ by the following reaction mechanism:

\begin{eqnarray}
	\text{S}^+ \xrightarrow{\text{CH}} \text{CS}^+ \xrightarrow{\text{H$_2$}} \text{HCS}^+      
  	\label{eq_1}
\end{eqnarray}

This mechanism is therefore very efficient because CH is rapidly formed by electronic recombination of CH$_2^+$ which is itself formed by the ion-neutral reaction between elemental species C$^+$ and H$_2$. Later, S is consumed by reactions with H$_3^+$, CH$_3$, OH and O$_2$, which efficiently produce SO causing the peak in its abundance at $3.6\times10^5$ years. The CS abundance follows closely the atomic sulphur one, first growing from numerous reactions, including notably the electronic recombination of HCS$^+$, C$_2$S$^+$ and HC$_3$S$^+$ and destruction of C$_2$S and by atomic carbon. It reaches a maximum between $10^4$ and $10^5$ years, becoming the second main reservoir of sulphur, containing up to 29\% of the initial sulphur. However, CS does not accumulate, in contrast to CO, as the hydrogenation of CS followed by DR of HCS$^+$ produce much more S + CH than H + CS \citep{Montaigne05}.\\"

It should be noted that we cannot reproduce the observed CS/HCS$^+$ ratio for TMC-1 of 26 \citep{Gratier16} at early time but the agreement become better, if not perfect, for longer age. Although HCS$^+$ and CS are linked through protonation reaction and DR, other reactions are involved in the chemistry of these species. Before $4.6\times10^3$ years, CS is mainly formed by DR of HCS$^+$ and HCS$^+$ is formed by reaction \ref{eq_1}. At longer time neutral reactions such as S + CH $\to$ CS + H plays also a role. Comparing our model results with the observed abundance of CS of $6.5\times10^{-9}$ \citep{Gratier16}, the problem is very likely an overestimation of the CS abundance, which can come from wrong branching ratios of the HCS$^+$ DR (we currently use the measured values from \citet{Montaigne05}), wrong observational estimations of CS (for example underestimation of CS due to saturation of CS lines), or an additional loss not producing HCS$^+$. A potentially new sink for CS may be the O + CS reaction. We currently use the rate from \citet{Lilenfeld77} determined in the 150-300 K range leading to a negligible rate at 10K. However, the work of \citet{Gonzalez96} may suggest a non negligible rate at low temperature.

\subsubsection{Main species on grains : HS ans H$_2$S} \label{HSandH_2S}

\begin{figure}
        \begin{center}
                \includegraphics[scale=0.14]{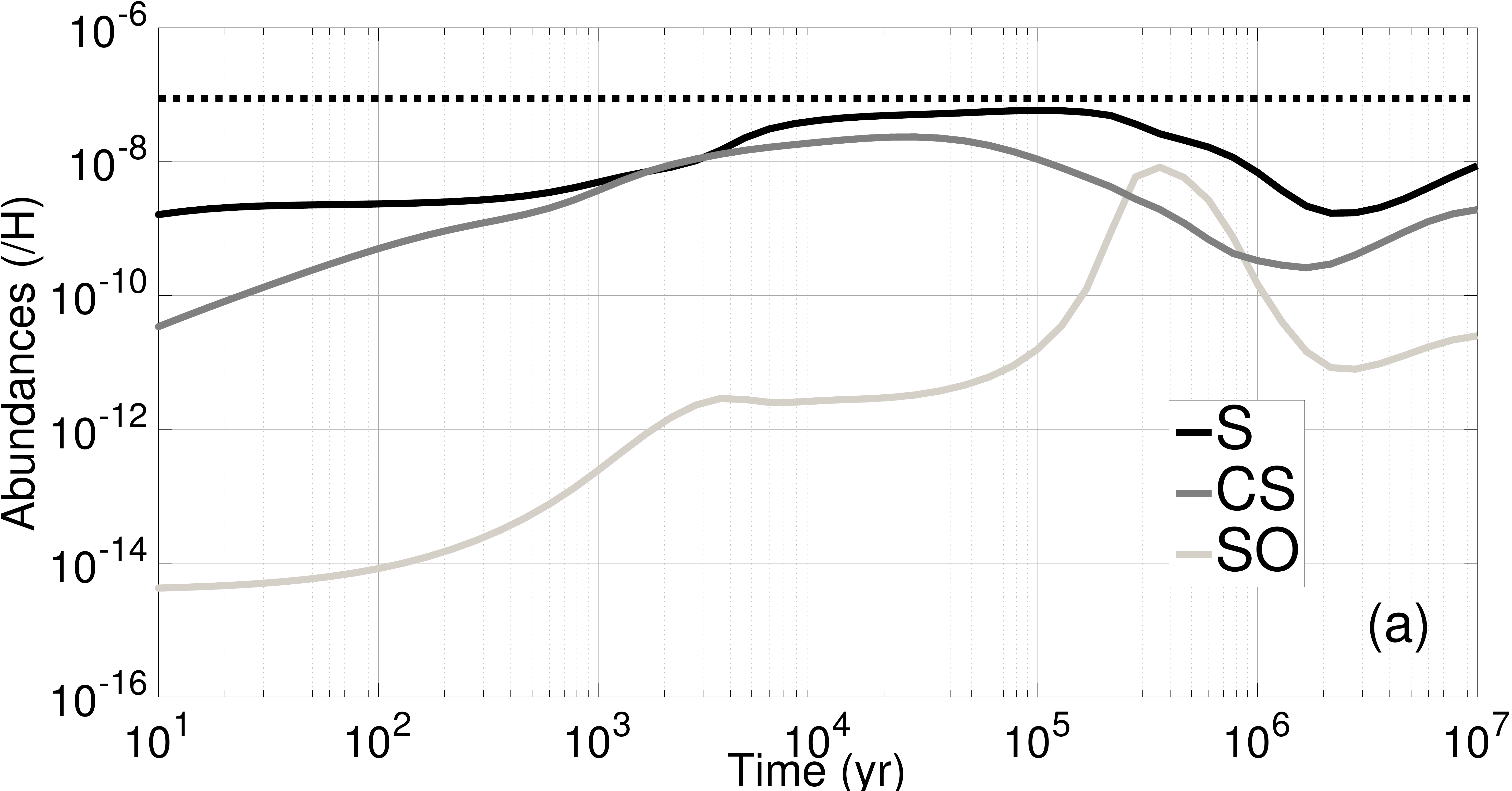}
                \includegraphics[scale=0.14]{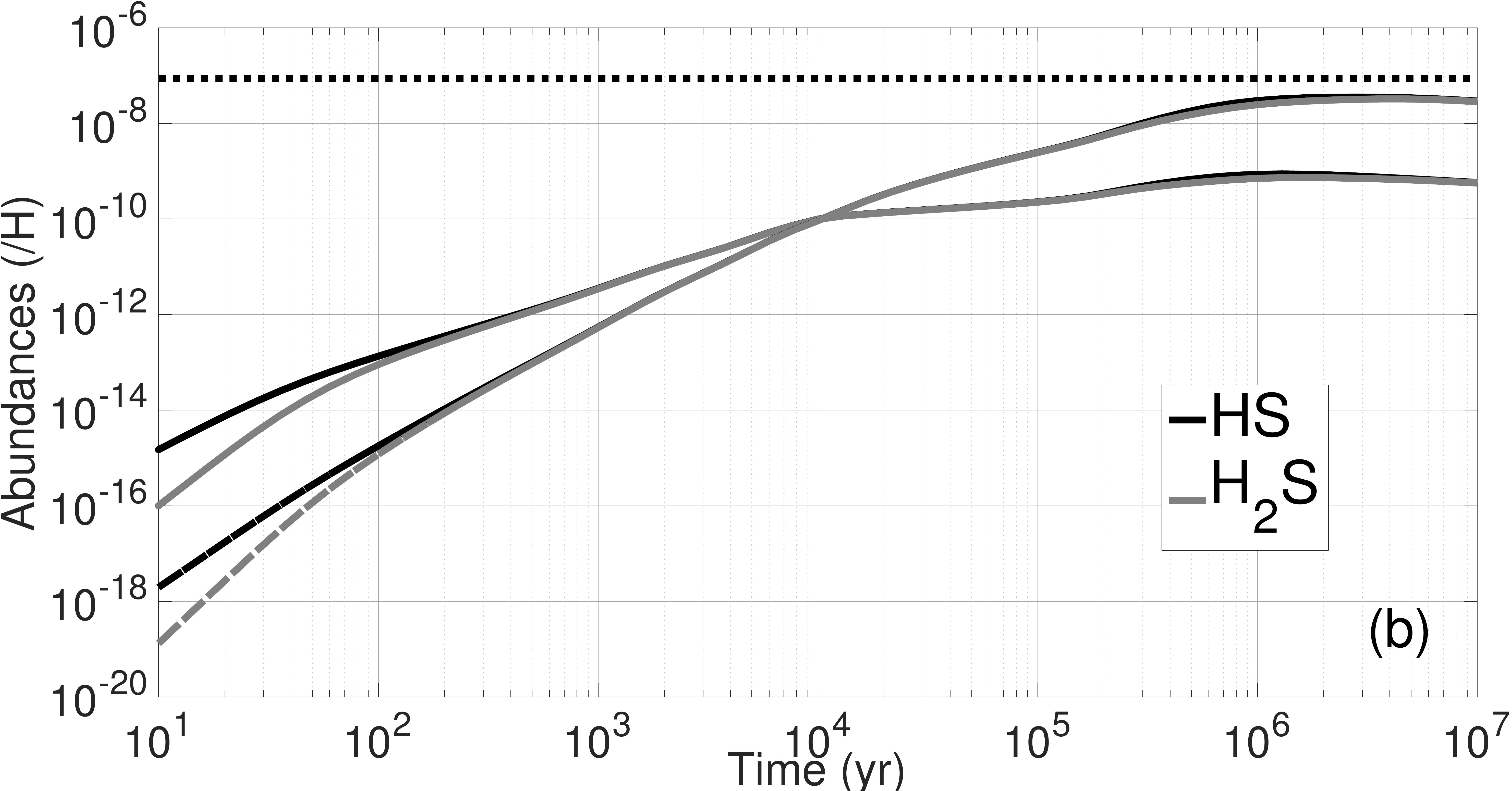}
                \caption{Abundances of main sulphur-bearing species relative to H as a function of time for dark cloud physical conditions : (a) in the gas phase, (b) on the grain surface (solid line) and bulk (dashed line). The dotted line represents the elemental abundance of sulphur (here, $X_{ini} = 8\times10^{-8}$).}
                \label{fig_1}
        \end{center}
\end{figure}

On the grains, it appears that the most abundant species are the hydrosulfide radical (HS) and hydrogen sulfide (H$_2$S). The abundances of those species, both on the surface (solid line) and in the bulk (dashed line), are depicted in Figure \ref{fig_1} (b). At times prior to 100 years, both species are formed on the surface by the successive hydrogenation of atomic sulphur physisorbed on the grains. Once the species are formed, they begin to sink into the bulk. After a hundred years, both species continue to form on the surface and eventually enter into a loop created by the two following reactions:

\begin{eqnarray}
   	\text{s-HS} + \text{s-H} &\to &\text{s-H$_2$S}\\
	\text{s-H$_2$S} + \text{s-H} &\to &\text{s-H$_2$}+\text{s-HS}
	\label{eq_2}
\end{eqnarray}

Reaction (\ref{eq_2}) has a barrier and should not be efficient under dark cloud physical conditions. However, it becomes efficient in our model because of the reaction-diffusion competition whose main effect is to increase the rates of reactions with activation barrier through tunneling \citep{Ruaud16}. Therefore, this cycle enabled by the reaction-diffusion competition and supplied in HS by the hydrogenation of S on the grains, keeps increasing at equal rates the abundances of the two species during the rest of the simulation. For this model, we have assumed a barrier of $Ea = 860$ K but there remains uncertainties in this value. The gas-phase reaction H + H$_2$S has been measured in a wide range of temperature from 190 K up to 2237 K \citep{Kurylo71, Peng99, Yoshimura92}. The rate constant shows strong non-Arhenius behavior at low temperature likely due to the importance of tunneling. \citet{Peng99} performed theoretical calculations of the barrier of the reaction equal to 1330 K at DFT level and 1930 K at QCISD(T) level. Considering these uncertainties, we have used the experimental value of \citet{Kurylo71} which is the one measured at the lower temperature (in the 190-464 K range). We also tested the higher values by \citet{Peng99} but it did not change significantly the results shown in this article. \\

As HS and H$_2$S keep on sinking into the bulk, their abundances in the latter eventually becomes higher than at the surface. At approximately $6\times10^5$ years and until $10^7$ years, HS and H$_2$S (in the bulk) are the main reservoirs of sulphur, sharing more than 80\% of the initial sulphur nearly equally divided between the two species, with a slight excess for HS. It should be noted that during this period of time, the 20 percent of the initial sulphur remaining is mostly divided between atomic S in the gas phase and in the bulk, as well as NS in the bulk.

\subsection{Abundances of the newly implemented S-bearing species}

Our enhanced sulphur network includes the addition of 45 new S-bearing species. Among those species, we now present the results of our calculations for HNCS, the species around which most of our new network has been built, and CH$_3$SH, which was recently detected in the envelope of the low mass protostar IRAS 16293-2422 \citep{Majumdar16}. 

\subsubsection{HNCS and HSCN}

\begin{figure}
        \begin{center}
                \includegraphics[scale=0.14]{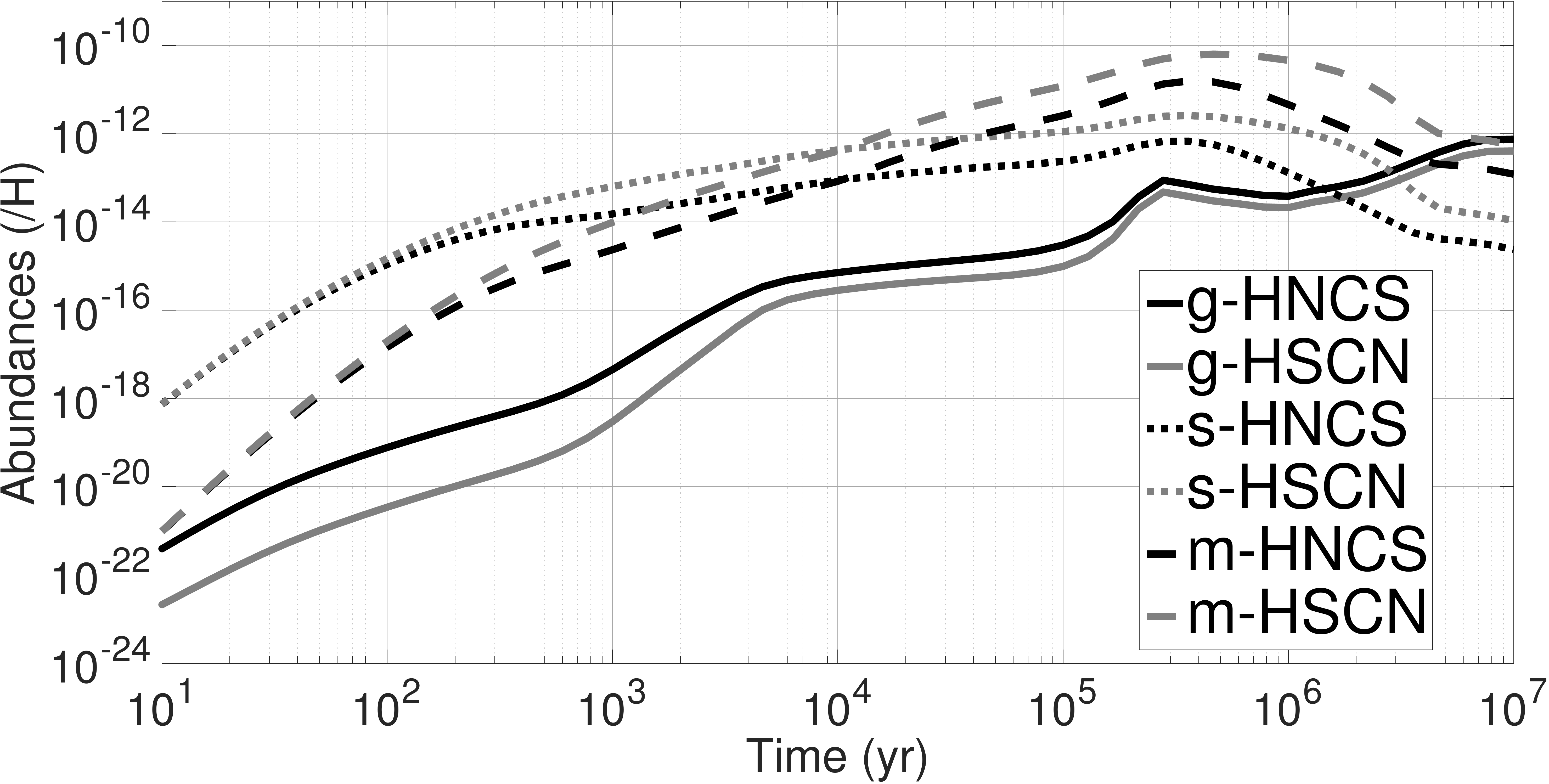}
                \caption{Abundances of HNCS (black) and HSCN (grey) relative to H as a function of time for dark cloud physical conditions in the gas phase (solid line), on the grain surface (dotted line) and in the grain bulk (dashed line).}
                \label{fig_6}
        \end{center}
\end{figure}

Isothiocyanic acid HNCS is the most stable among possible CHNS isomers. It has been very well studied since its first detection in Sgr B2 by \citet{Frerking79}, and has since been detected in the dark cloud TMC-1 (CP) \citep{Adande10}. In contrast, thiocyanic acid HSCN is highly unstable and is about a factor of 3 less abundant in Sgr B2 than HNCS \citep{Halfen09}, and has been also detected in TMC-1 (CP) with a similar abundance to HNCS \citep{Adande10}. Figure \ref{fig_6} shows the time evolution of their abundances in the three phases of the model (in the gas phase, on the grain surface and in the grain bulk) for dark clouds conditions. It should be noted that our network for HNCS and HSCN is an improved version of the one proposed in \cite{Gronowski14}, which itself is a revised version of the network proposed in \cite{Adande10}. It appears from our simulations that both species are essentially produced at the surface of the grains where their main formation reaction is: 

\begin{eqnarray}
   	\text{s-N} + \text{s-HCS} &\to& \text{s-HNCS or s-HSCN}
	\label{eq_12}
\end{eqnarray}

HNCS and HSCN abundances on the grain surface are also regulated by hydrogenation loops similar to the one described in section \ref{HSandH_2S} : for HNCS with HNCSH and NH$_2$CS, and for HSCN with HNCSH and HSCHN. It should be noted that the hydrogenation of HSCN is calculated to be much less efficient than the hydrogenation of HNCS (see appendix A). At times $>2.8\times10^5$ years, as less and less atomic N is available, abundances of both species decrease on the surface. For HNCS it happens through the chemical desorption of the products of its hydrogenation, HNCSH and NH2CS. For HSCN it happens through its successive hydrogenations. Moreover, as the chemistry of both species is not efficient in the bulk, its abundance increase in this phase is primarily due to sinking from the surface.\\

In the gas phase, HNCS and HSCN abundances increase efficiently from chemical desorption from the grains following reaction (\ref{eq_12}), as well as from the following reaction mechanism:

\begin{eqnarray}
	\text{HCS}^+ + \text{NH}_2 \to \text{H} + \text{H$_2$NCS}^+ \xrightarrow{\text{e-}}  \text{HNCS/HSCN} + \text{H} 
	\label{eq_12b}
\end{eqnarray}

Consequently, after $2.8\times10^5$ years, as reaction (\ref{eq_12}) is less efficient because there is less N available, reaction (\ref{eq_12b}) then becomes the main formation pathway of HNCS and HSCN in the gas phase. They are also predicted to be efficiently destroyed by atomic C via: 

\begin{eqnarray}
   	\text{C} + \text{HNCS} &\to &\text{HNC} + \text{CS}\\
                                             &\to &\text{HCN} + \text{CS}\\
        \text{C} + \text{HSCN} &\to &\text{HCN} + \text{CS}
	\label{eq_13}
\end{eqnarray}

The depletion of atomic C on grains after $10^5$ years then explains the increase in the gas phase abundance gradients for both species at this time.

\subsubsection{CH$_3$SH}

\begin{figure}
        \begin{center}
                \includegraphics[scale=0.14]{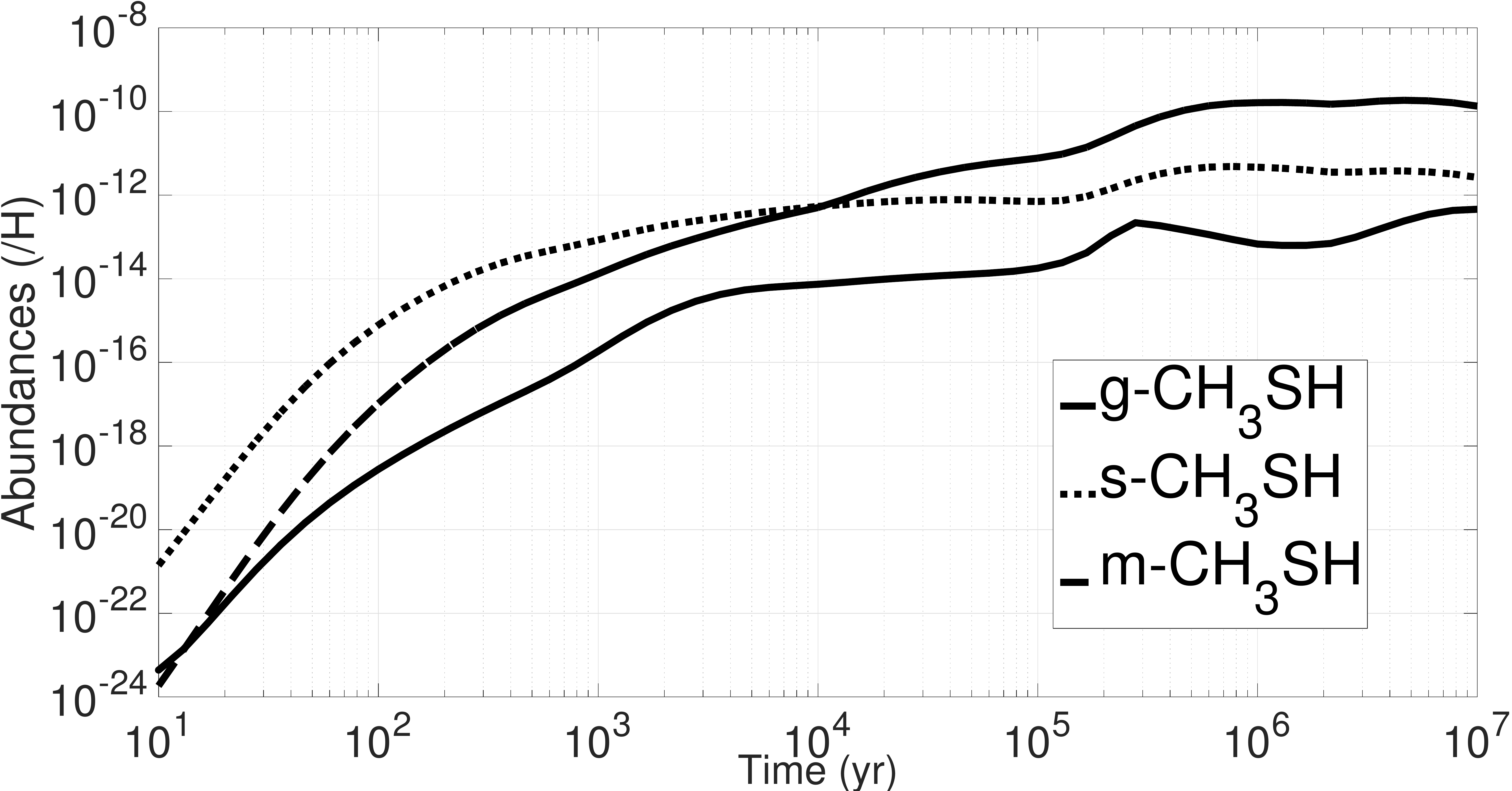}
                \caption{Abundance of CH$_3$SH relative to H as a function of time for dark cloud physical conditions in the gas phase (solid line), on the grain surface (dotted line) and in the grain bulk (dashed line)}
                \label{fig_7}
        \end{center}
\end{figure}

Numerous previous detections of methyl mercaptan CH$_3$SH in hot cores such as Sgr B2 \citep{Linke79}, G327.3-0.6 \citep{Gibb00} and Orion KL \citep{Kolesnikov14} suggest that this species initially forms in the ices and then evaporates in hot cores. More recently, \citet{Majumdar16} presented the detection of CH$_3$SH in IRAS 16293-2422 and proposed an associated chemical network to explain its observed abundance. This work is included in our enhanced sulphur network and has been completed, notably regarding grain chemistry. Figure \ref{fig_7} shows the time evolution of the CH$_3$SH abundance. It appears from our results that CH$_3$SH is mainly formed on the grain surface and is released into the gas phase by chemical desorption in dark clouds conditions. The main formation processes on the grains are: 

\begin{eqnarray}
   	\text{s-H} + \text{s-CH$_3$S} &\to& \text{s-CH$_3$SH}\\
	\text{s-H} + \text{s-CH$_2$SH} &\to& \text{s-CH$_3$SH}
	\label{eq_14}
\end{eqnarray}

Moreover, similarly to HNCS, it seems that CH$_3$SH chemistry in the grain bulk is relatively inert, which throughout the simulation explains its accumulation in this phase from sinking from the grain surface.\\

As said above, the gas phase chemistry of CH$_3$SH is essentially ruled by its chemical desorption from the grain surface during the entire simulation. As for HNCS, it should be noted that it is consumed by atomic C at times $<2.8\times10^5$ years following:

 \begin{eqnarray}
   	\text{C} + \text{CH$_3$SH} &\to &\text{CH$_3$} + \text{HCS}
	\label{eq_15}
\end{eqnarray}

However after this time, in contrast to HNCS, gas phase CH$_3$SH is still produced from grain surface chemistry and chemical desorption, which can be explained by the fact that the CS abundance in the gas phase remains high, and gradually depletes to form CH$_3$SH through successive hydrogenations. Moreover, as the hydrogen atom abundance remains relatively constant on the grain surface during the simulation, due to cosmic-ray interactions with H$_2$, CH$_3$SH is efficiently produced in contrast to HNCS and HSCN as nitrogen atomic strongly decreases.

\subsection{Comparison with the previous version of the network} \label{NetComp}

We now highlight the differences between the nominal model (hereafter Model A) and Model B, which takes into account our enhanced sulphur network. It should be noted that Model A already includes the reactions between HS and H$_2$S studied in section \ref{HSandH_2S}, as well as the reaction-diffusion competition mechanism, two key parameters for sulphur chemistry. As our calculations show that numerous S-bearing species are impacted by the changes we made to the network, we present here only a selection of these species (C$_3$S, H$_2$CS, HCS, H$_2$S, HS and SO). We selected these molecules because they present a difference between Models A and B of more than one order of magnitude in the time period between $10^5$ and $10^6$ years and because their abundances rise above $10^{-12}$ during this period. Indeed, the estimated ages of well-studied dark clouds such as TMC-1 or L134N are thought to correspond to this time period \citep[see discussion in][]{Agundez13}. Hence, an abundance of $10^{-12}$ is considered as the detection limit for species in the gas-phase.\\

\subsubsection{C$_3$S}

\begin{figure}
        \begin{center}
                \includegraphics[scale=0.14]{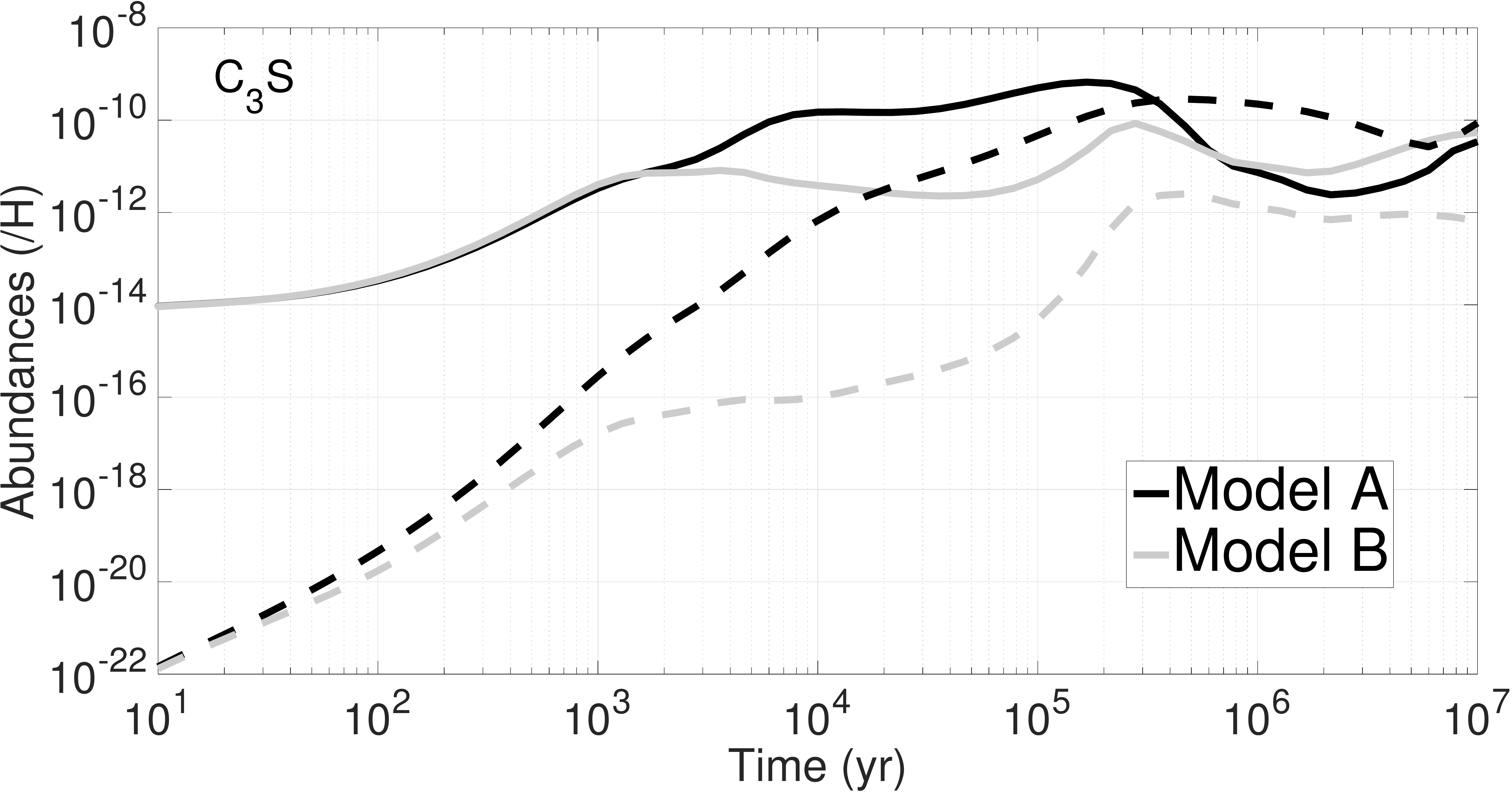}
                \caption{Comparison of the abundances of C$_3$S relative to H as a function of time for dark cloud physical conditions in the gas phase (solid line) and in the grain bulk (long-dashed line).}
                \label{fig_2}
        \end{center}
\end{figure}

Figure \ref{fig_2}  shows the comparison of the abundances of C$_3$S calculated for models A and B, both in the gas phase and in the grain bulk. It can be seen that our new network induces differences for this species that can reach up to two orders of magnitude in the gas phase and four orders of magnitude in the grain bulk at times around 10$^5$ years. In both models, C$_3$S is mainly formed in the gas phase by the reaction: 

\begin{eqnarray}
	\text{HC$_3$S}^+ + \text{e}^- \to \text{H}+\text{C$_3$S}
  	\label{eq_3}
\end{eqnarray}

HC$_3$S$^+$ being efficiently produced through reactions of atomic sulphur with c,l-C$_3$H$_2^+$ and c,l-C$_3$H$_3^+$. We thus consider that the decrease of its abundance in the grain bulk in Model B (as compared to Model A) is a direct result of its decrease in the gas phase. This decrease is due to the following added reaction:
 
 \begin{eqnarray}
	\text{C} + \text{C$_3$S} \to \text{C$_3$}+\text{CS}
  	\label{eq_4}
\end{eqnarray}

This reaction becomes efficient in Model B at times $>10^3$ years when the abundance of atomic C in the gas phase reaches $10^{-4}$. Finally, after $10^4$ years, while in Model A the abundance of C$_3$S continues to increase because of reaction (\ref{eq_3}), in Model B the small peak at $2.8\times10^5$ years is due to the reaction: 

 \begin{eqnarray}
	\text{O} + \text{HC$_3$S} \to \text{C$_3$S}+\text{OH}
  	\label{eq_5}
\end{eqnarray}

Reaction (\ref{eq_5}) becomes more efficient than reaction (\ref{eq_4}) at this time because depletion is more efficient for atomic carbon ($\sim 10^5$ years) than for atomic oxygen ($\sim 8\times10^5$ years). It should be noted that branching ratios for the DR of HC$_3$S$^+$ are unknown and our network may overestimate C$_3$S production leading to an overestimation of its abundance.\\

\subsubsection{H$_2$CS}

\begin{figure}
        \begin{center}
                \includegraphics[scale=0.14]{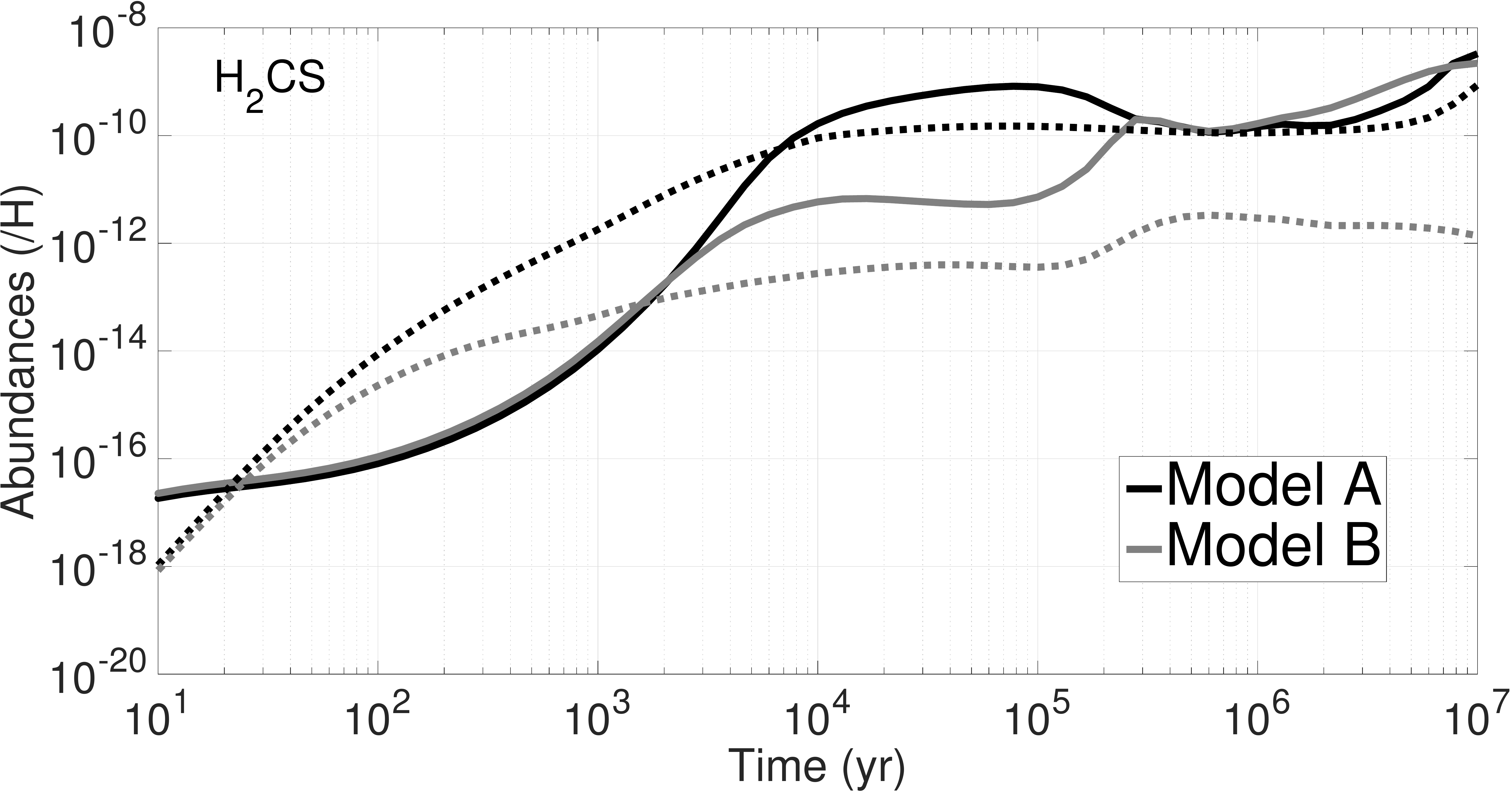}
                \caption{Comparison of the abundances of H$_2$CS relative to H as a function of time for dark cloud physical conditions in the gas phase (solid line) and on the grain surface (dotted line).}
                \label{fig_3}
        \end{center}
\end{figure}

The abundance of H$_2$CS in the gas phase presents a similar change in its evolution between Model A and B (see Figure \ref{fig_3}). Its main reaction of formation in both models is the electronic recombination of H$_3$CS$^+$: 

\begin{eqnarray}
	\text{H$_3$CS}^+ + \text{e}^- \to \text{H}+\text{H$_2$CS}
  	\label{eq_6}
\end{eqnarray}

H$_3$CS$^+$ being efficiently produced through S$^+$ + CH$_4$ reaction. H$_2$CS is less efficiently formed in Model B after $10^3$ years because, like C$_3$S, it is consumed by atomic C via the following two reactions: 

\begin{eqnarray}
   	\text{C} + \text{H$_2$CS} &\to &\text{H}+\text{HC$_2$S}\\
	\text{C} + \text{H$_2$CS} &\to &\text{CH$_2$}+\text{CS}
	\label{eq_8}
\end{eqnarray}

These reactions induce differences in the H$_2$CS abundance of more than two orders of magnitude between $10^4$ and $10^5$ years. As atomic carbon is depleted onto the grains, these reactions become less efficient after $10^5$ years and the abundance of H$_2$CS increases again through reaction (\ref{eq_6}).\\

In contrast to C$_3$S, H$_2$CS is also efficiently formed on the grain surface in both models by the hydrogenation of HCS. However, in Model B, two hydrogenation processes of H$_2$CS have been added, which cause its abundance at the grain surface to drop compared to Model A by more than two orders of magnitude. Those processes are: 

\begin{eqnarray}
   	\text{s-H$_2$CS} + \text{s-H}   &\to &\text{s-CH$_2$SH}\\
	                           		        &\to &\text{s-CH$_3$S}
	\label{eq_9}
\end{eqnarray}

which also lead to the formation of CH$_3$SH. Moreover, as already noted, the branching ratios for the DR of H$_3$CS$^+$ are unknown and may lead mainly to breaking of the C-S bond which will strongly limit the H$_2$CS abundance.

\subsubsection{HCS} \label{HCS}

\begin{figure}
        \begin{center}
                \includegraphics[scale=0.14]{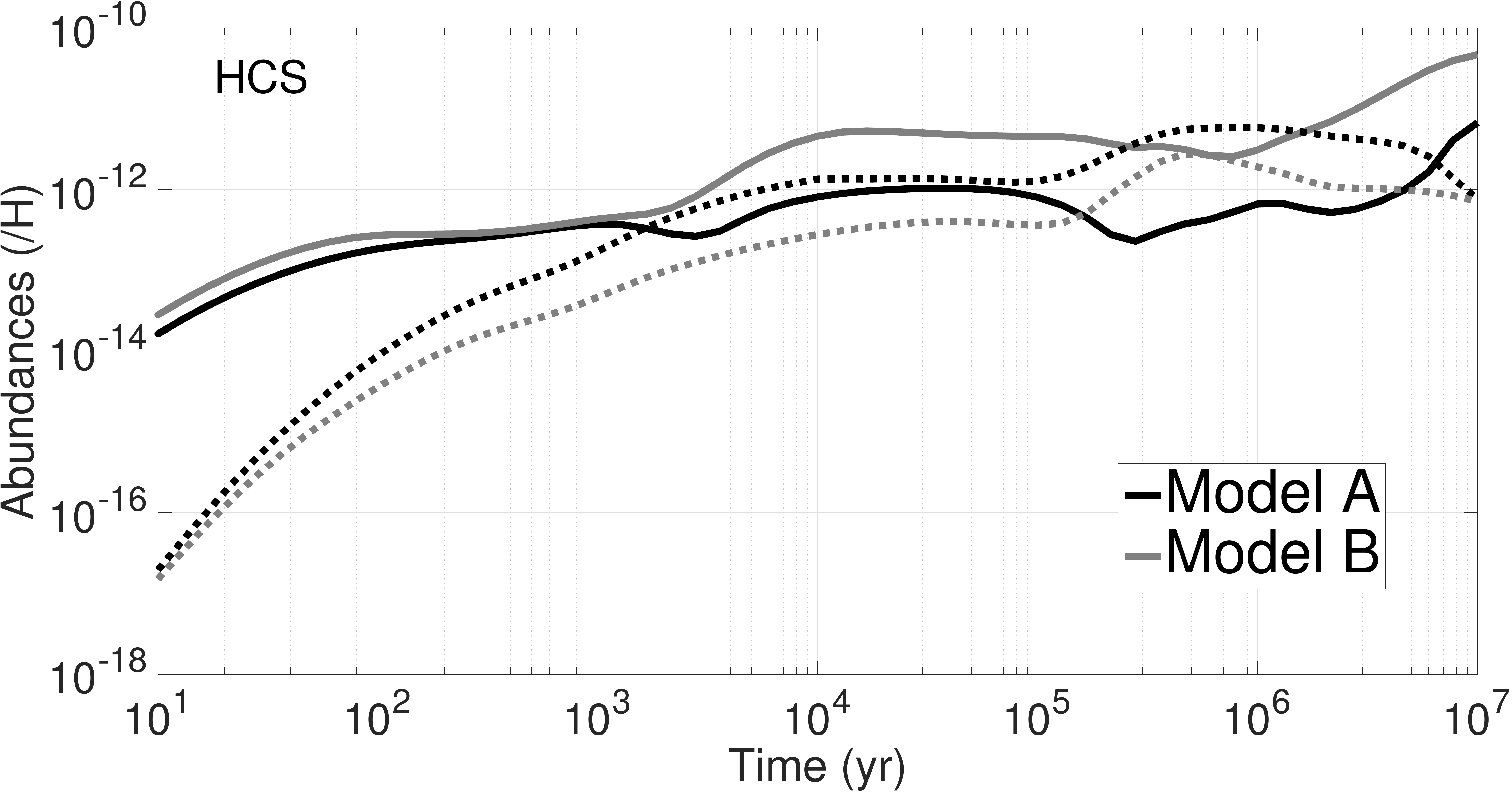}
                \caption{Same as Figure \ref{fig_3} but for HCS.}
                \label{fig_4}
        \end{center}
\end{figure}

Figure \ref{fig_4}  shows the time evolution of the HCS abundance in the gas phase in both Models A and B. We can see in this figure that the overall effect of our enhanced network on this species is to increase its abundance in the gas phase above the detection level of $10^{-12}$. This potential observability of HCS in dark clouds will be discussed later. In Model A, HCS is mainly formed by the two following reactions: 

\begin{eqnarray}
   	\text{S} + \text{CH$_2$} &\to &\text{H}+\text{HCS}\\
	\text{H$_2$CS$^+$} + \text{e}^- &\to &\text{H}+\text{HCS}
	\label{eq_10}
\end{eqnarray}

However, in Model B two new reactions show a higher HCS production efficiency. Those reactions are:

\begin{eqnarray}
	\text{H$_3$CS$^+$} + \text{e}^- &\to &\text{H}+\text{H}+\text{HCS}\\
	\text{C} + \text{H$_2$S} &\to &\text{H}+\text{HCS}
	\label{eq_10a}
\end{eqnarray}

It should be noted that HCS is also formed at the grain surface by the hydrogenation of CS, and that its abundance in this phase is decreased in Model B because of numerous added reactions with atomic O and N, the latter forming the newly implemented species HNCS on the grains.

\subsubsection{HS, H$_2$S and SO}

\begin{figure}
        \begin{center}
                \includegraphics[scale=0.14]{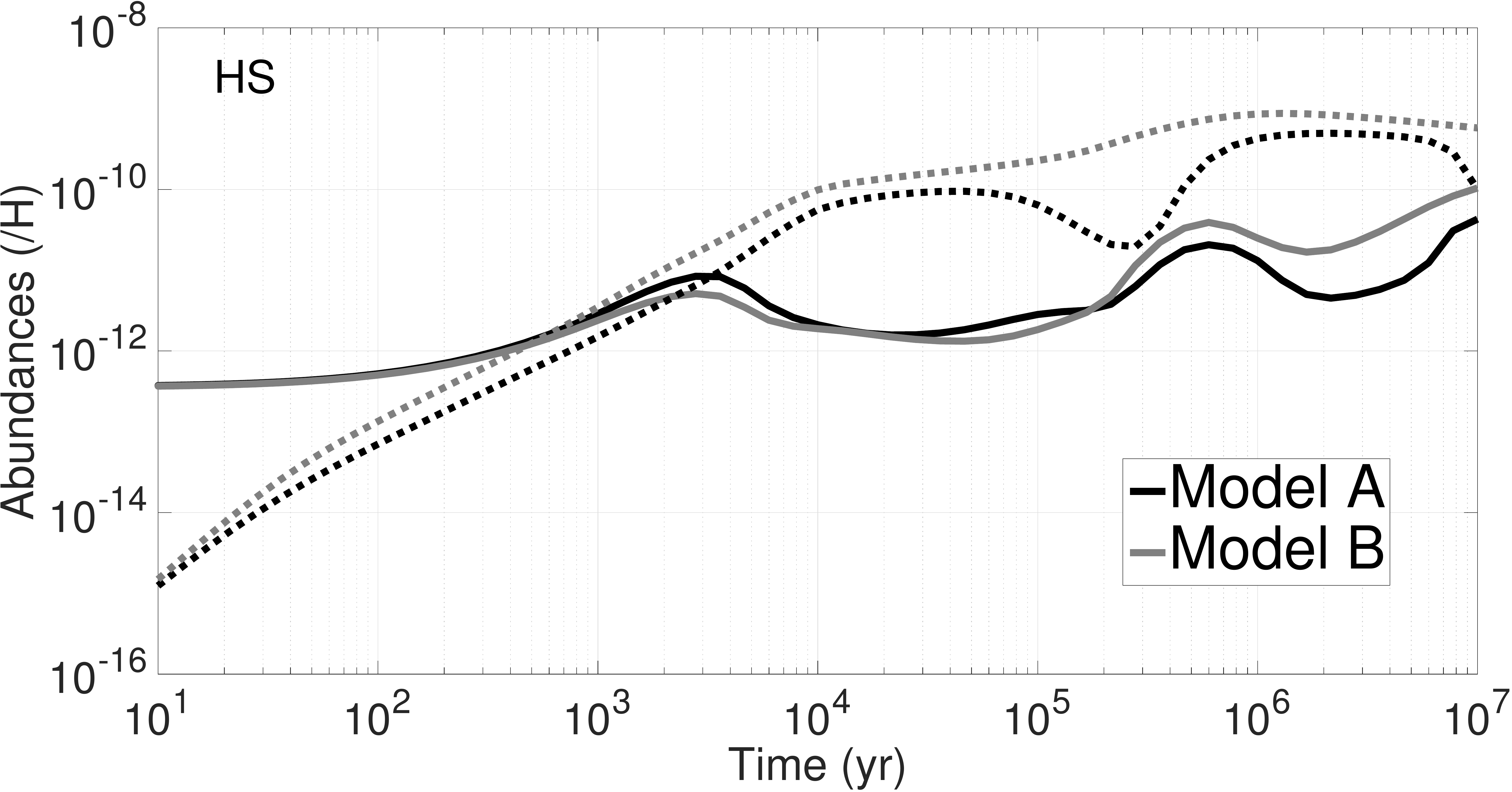}
                \includegraphics[scale=0.14]{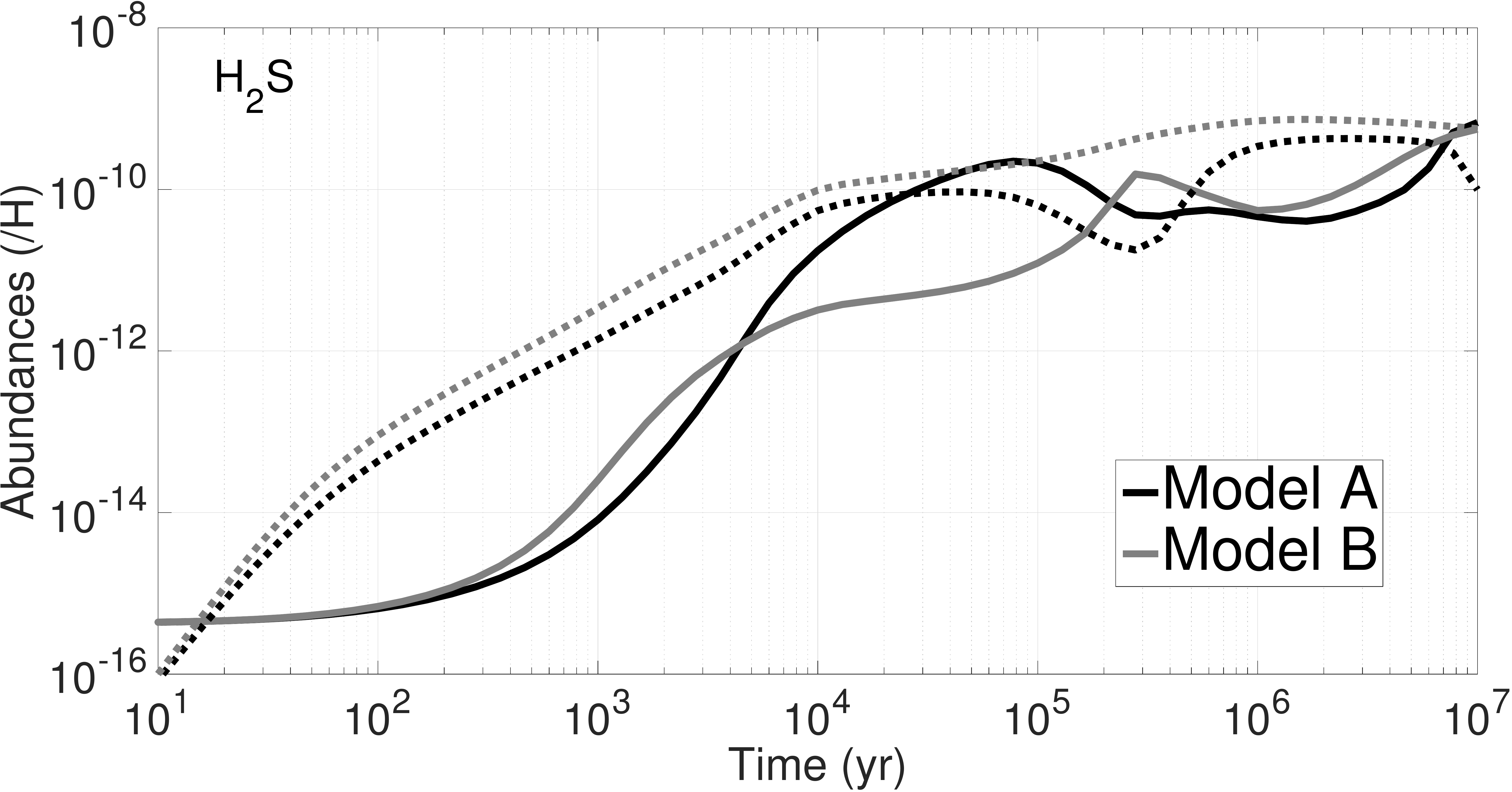}
                \includegraphics[scale=0.14]{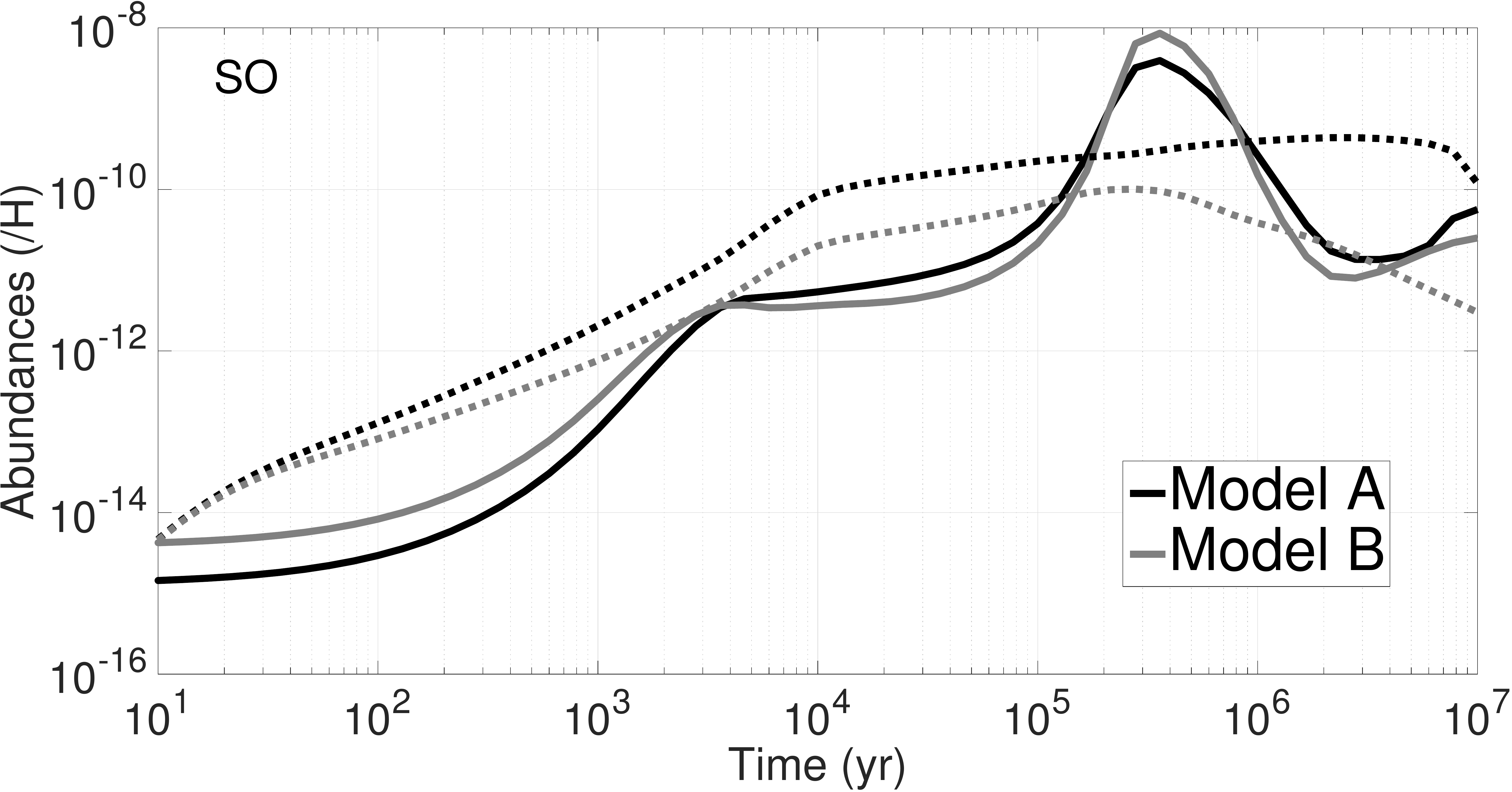}
                \caption{Comparison of the abundances of HS, H$_2$S and SO, relative to H as a function of time for dark cloud physical conditions in the gas phase (solid line) and on the grain surface (dotted line).}
                \label{fig_5}
        \end{center}
\end{figure}

We choose to describe the changes brought by our new model to HS, H$_2$S and SO abundances together, notably on the grain surface, because they are correlated.\\ 

On the one hand, in the gas phase, among those three species, the only one that presents a significant difference in Model B (compared to Model A) is H$_2$S (see Figure \ref{fig_5}). This difference (that reaches more than one order of magnitude) appears between $5\times10^3$ and $3\times10^5$ years and is mainly due to the fact that reaction (\ref{eq_10a}) slows down the increase of H$_2$S during this period of time. Besides this change, gas phase H$_2$S is efficiently formed in both models on the grain surface by the hydrogenation of HS (reaction (\ref{eq_2})) enabled by the reaction-diffusion competition, followed by desorption.\\

On the other hand, our modifications of the network bring about a new reaction mechanism on the grain surface which causes several changes to the abundances of HS, H$_2$S and SO:

\begin{eqnarray}
	\text{s-S}  &\xrightarrow{\text{s-H}} &\text{s-HS}\\
	                &\xrightarrow{\text{s-O}} &\text{s-SO} \xrightarrow{\text{s-H}}  \text{s-HSO} \xrightarrow{\text{s-H}} \text{s-HS} 
	\label{eq_11}
\end{eqnarray}

From Model A to Model B, the main effects of this reaction mechanism are: 

\begin{itemize}
\item to globally increase the grain surface abundance of HS because of the added path (\ref{eq_11}),\\
\item to globally increase the grain surface abundance of H$_2$S because of the loop described in section \ref{HSandH_2S}, and hence its abundance in the gas phase (except during the time period described above),\\
\item to globally decrease the grain surface abundance of SO which is, in our enhanced network, successively hydrogenated to form HSO then HS.
\end{itemize}

\section{Comparison with observations} \label{compobs}

In this section we compare our modeling results with observations of sulphur bearing species in dense molecular clouds. We use the method of the so-called distance of disagreement described in \cite{Wakelam06} and which is computed as follows: 

\begin{eqnarray}
	D(t) = \frac{1}{N_{obs}}\sum_{i}\mid\log[n(X)_{obs,i}]-\log[n(X)_i(t)]\mid
	\label{eq_16}
\end{eqnarray}

where $n(X)_{obs,i}$ and $n(X)_i(t)$ are the observed abundances (relative to H) of the $i$-th species and the modeled ones at time $t$, respectively. $N_{obs}$ is the total number of observed species considered. The minimum of function $D$ (hereafter noted $D_{min}$) is then the minimum average difference (in magnitude) between modeled and observed abundances. We call the time for which this minimum is obtained the "best fit" time (hereafter noted $t_{BF}$) which consequently is the estimated age of the observed object according to the model.\\

We detail the observed abundances we use for S-bearing species in table \ref{tab_4}, and for other species we use Table 4 of \citet{Agundez13} and reference therein.

\begin{table}
\caption{Observed abundances of S-bearing species in TMC-1 (CP) . *$a(b)$ stands for $a\times10^b$}
	\begin{center}
		\begin{tabular}{l r r}
		\hline
		\hline
   		Species & $n_i/n_H$* & References \\
   		\hline
		OCS            & 1.1(-9)        &1 \\
   		NS              & 4.0(-10)     & 2 \\
		CS              & 6.5(-9)       &  3 \\
		HCS$^+$    & 2.5(-10)    &  3 \\
		H$_2$CS    & 2.6(-9)     &  3 \\
		C$_2$S      & 5.0(-9)      &  3 \\
		C$_3$S      & 7.0(-10)    &  3 \\
		SO 		  & 1.0(-8)    &  4 \\  		
		SO$_2$	  & 1.5(-10)  &  5 \\
		HNCS  	  & 4.15(-12)  &  6 \\
		HSCN  	  & 6.3(-12)  &  6 \\
		\hline
 		\end{tabular}
	\end{center}
	\medskip{(1) \citet{Matthews87}, (2) \citet{McGonagle94}, (3) \citet{Gratier16}, (4) \citet{Lique06}, (5) \citet{Cernicharo11}, (6) \citet{Adande10}}
  	\label{tab_4}
\end{table}

\subsection{Comparison with Models A and B}

In order to validate our enhanced sulphur network we compare the results of both Models A and B with the observations of the well-studied TMC-1 (CP) dark cloud for which numerous observational constraints are available. Indeed, more than 60 gas phase species have been detected in this source, and upper limits on the abundance of 7 more are available \citep[see a review of these species in][]{Agundez13,Adande10}. For the calculation of the corresponding distance of disagreement for both models, we take into account 58 of the 64 detected species because 6 of them are not implemented in one or either of the two networks (nominal and enhanced). Using equation (\ref{eq_16}), we find a similar level of agreement between Model A and B, with however a slightly better agreement with observations for Model B with $D_{min} = 0.766$ (against $D_{min} = 0.908$ for Model A), giving a "best fit" time of $2.8\times10^5$ years, same as Model A. At this time, and assuming that the observed abundances were reproduced by the model when the difference between the two was smaller than one order of magnitude, the fractions of reproduced molecules are 64\% and 67\% respectively for Model A and Model B. A summary of these results is shown in Table \ref{tab_2}.\\

For both models, 5 of the 9 S-bearing species taken into account in the comparison are reproduced : OCS, CS, C$_3$S, SO and SO$_2$, and Model B gives a slightly better agreement than Model A for these species. The similarity of the results comes from the fact that Model A includes the reaction-diffusion competition as well as the hydrogenation loop described in section \ref{HSandH_2S}, both key parameters for sulphur chemistry. Moreover, considering only the 58 species taken into account in Model A, Model B reproduces only two more species than Model A which are carbon chains. However, the goal of this article is to evaluate the efficiency of our enhanced network (Model B) to reproduce S-bearing species abundances, which should be done by considering all the species it takes into account. Therefore we hereafter choose to exclude Model A from our study and focus only on Model B.\\

\begin{table}
\caption{Results of the comparison of Models A and B with observations in the dark cloud TMC-1 (CP)}
	\begin{center}
		\begin{tabular}{c c c c}
		\hline
   		Model & $D_{min}$ & $t_{BF}$ (yr) & Fraction of reproduced molecules\\
   		\hline
		\hline
		A        & $0.908$     & $2.8\times10^5$  &  64\%\\
		B        & $0.766$     & $2.8\times10^5$  &  67\%\\
		\hline
 		\end{tabular}
	\end{center}
  	\label{tab_2}
\end{table}

\subsection{Variation of the elemental sulphur abundance}

In order to assess the issue of sulphur depletion in chemical simulations of dark clouds with our enhanced network, we run the model for 4 different values of the elemental sulphur abundance and compare the results with observations to determine which one leads to better agreement with the observed species abundances. These values vary from the so-called depleted one ($8\times10^{-8}$), which is commonly used by chemical models to reproduce observations and which we used previously in this paper, to the cosmic one ($1.5\times10^{-5}$), which is considered to be the real elemental abundance of sulphur for dark cloud formation. The different values of the sulphur elemental abundances used and the labels of their associated models are summarized in Table \ref{tab_3}. Apart from sulphur, the other species elemental abundances we use are those reported in Table \ref{tab_1}.\\

\begin{table}
\caption{Set of elemental sulphur abundances. *$a(b)$ stands for $a\times10^b$}
	\begin{center}
		\begin{tabular}{c c}
		\hline
   		Model & $n(S)_{ini}$*\\
   		\hline
		\hline
		1        & $8(-8)$\\
		2        & $5(-7)$\\
		3        & $5(-6)$\\
		4        & $1.5(-5)$\\
		\hline
 		\end{tabular}
	\end{center}
  	\label{tab_3}
\end{table}

In the following, we first compare for each model our calculated abundances of gas-phase species with those determined from the well-studied dark cloud TMC-1 (CP). The only sulphur-bearing species observed on ice being OCS and SO$_2$ in the molecular cloud surrounding the deeply embedded protostar W33A, we then compare the modeled abundances of these species on the grains with the observed ones.

\subsubsection{Comparison with all observed gas-phase species in TMC-1 (CP)}

\begin{figure}
        \begin{center}
                \includegraphics[scale=0.14]{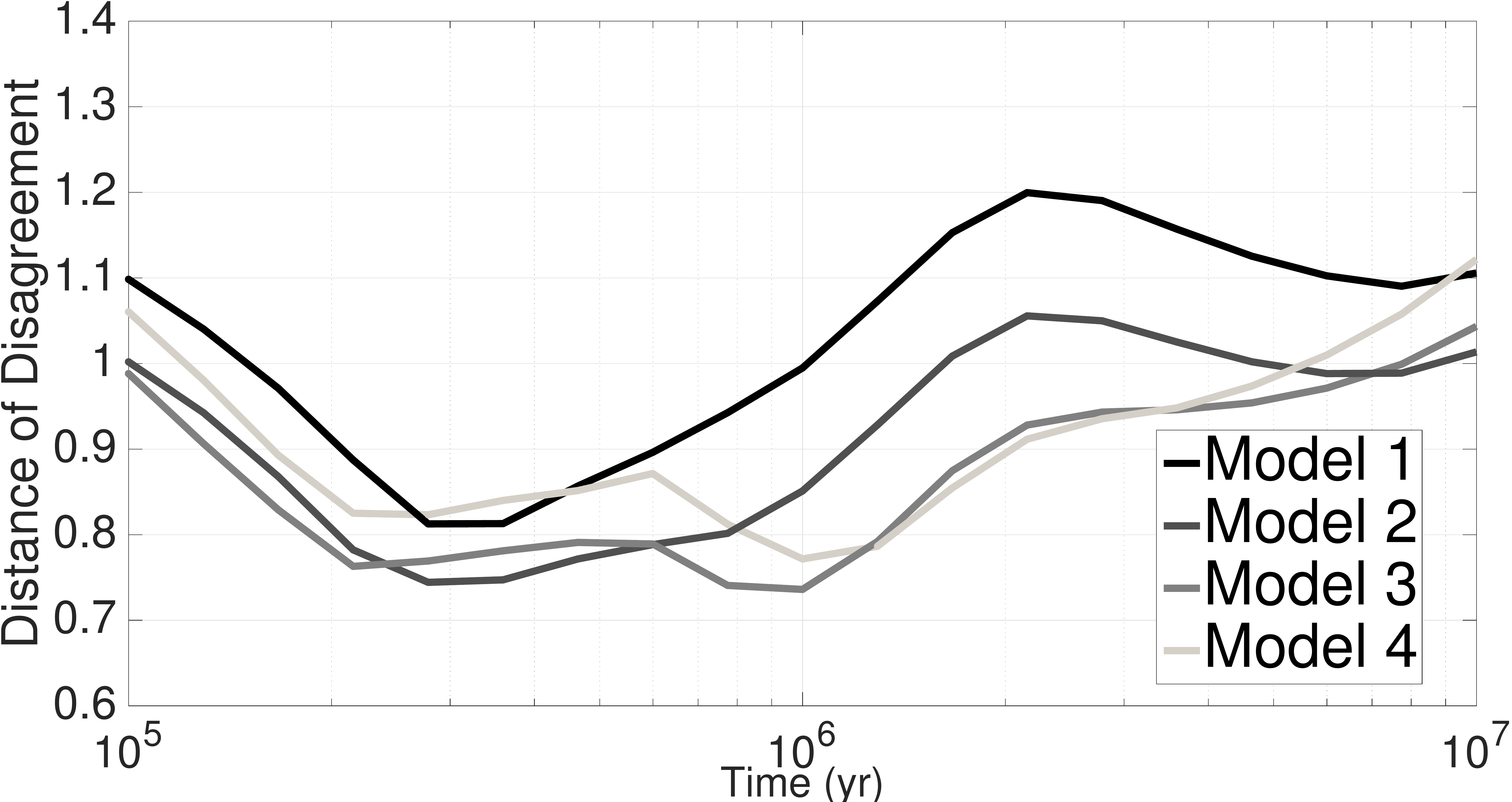}
                \caption{Comparison of the distances of disagreement for Models 1, 2, 3 and 4 considering 61 detected species in TMC-1 (CP).}
                \label{fig_8}
        \end{center}
\end{figure}

To assess the overall agreement of our enhanced network with observations, we compute the distance of disagreement for Models 1, 2, 3 and 4 with the 61 observed species that are taken into account in this network. Figure \ref{fig_8} displays these distances of disagreement between $10^5$ and $10^7$ years, the time period when the minimum values are reached. On the one hand, it appears that Model 3 is the best model in that it presents the distance of disagreement with the lowest minimum of $D_3(t_{BF})=0.736$ with $t_{BF}=10^6$ years which is still a reasonable estimated age for TMC-1 (CP). Moreover at that time, 44 of the 61 species considered (72\%) are reproduced by the model within a factor of 10. This result means that the elemental sulphur abundance which our new model theoretically needs to best reproduce the observations is around $X(S)_{ini}=5\times10^{-6}$. Model 4 also reproduce 44 of the 61 species at its 'best fit' time but its minima of distance of disagreement is slightly higher than Model 3 (0.772).\\

On the other hand, Figure \ref{fig_8} shows that the maximum difference between the distances of disagreement of the four Models is lower than 0.3. Hence, for a given time, the average of the differences of abundances between two Models (for a same species) is lower than a factor of 2. The distance of disagreement is thus not very sensitive to the elemental sulphur abundance. This result can be explained by the fact that among the 61 observed species we consider for the calculations, only 11 of them are S-bearing species.

\subsubsection{Comparison with observed S-bearing gas-phase species in TMC-1 (CP)}

\begin{figure}
        \begin{center}
                \includegraphics[scale=0.14]{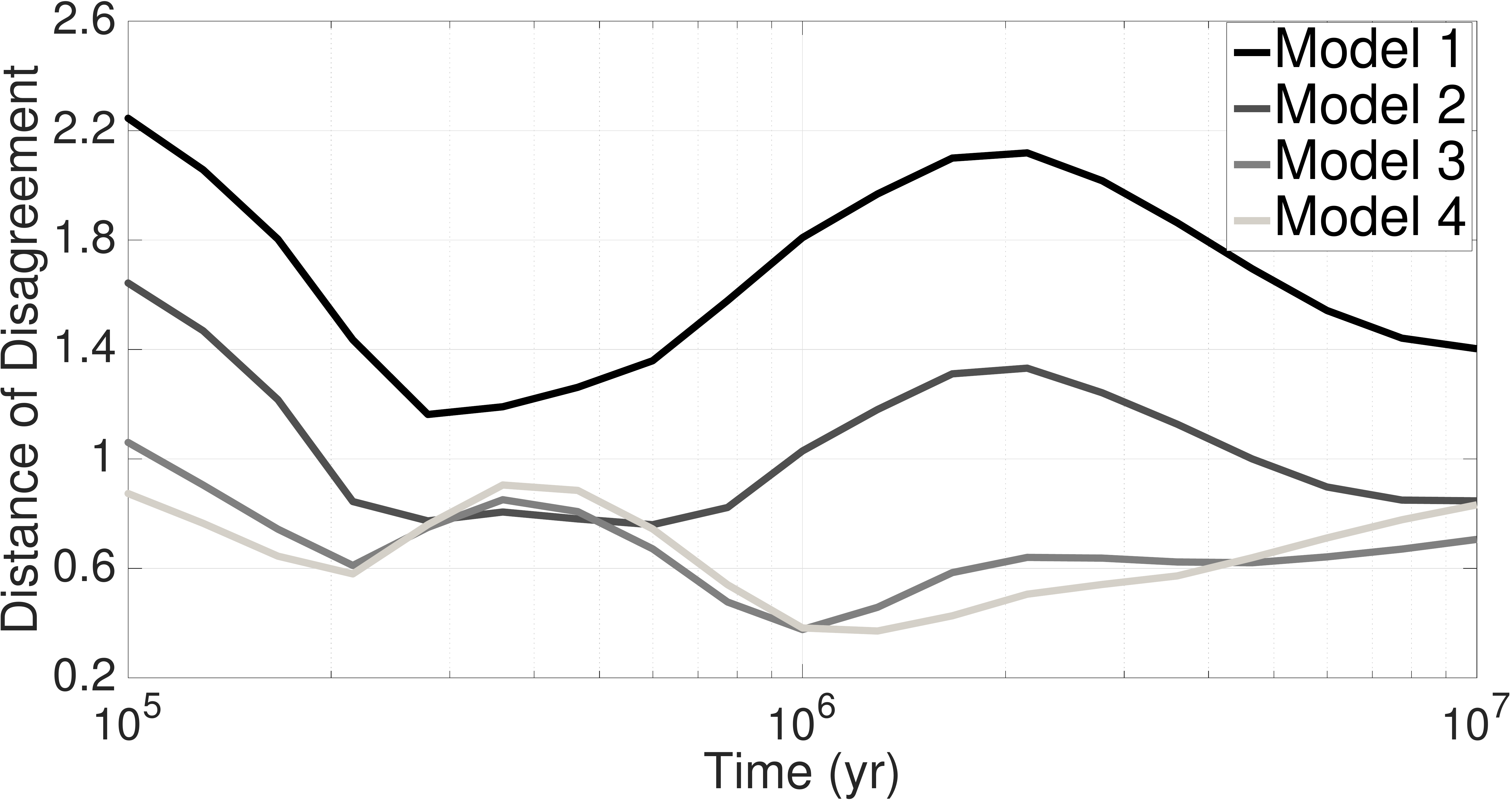} 
                \caption{Comparison of the distances of disagreement for Models 1, 2, 3 and 4 considering only the 11 detected S-bearing species in TMC-1 (CP).}
                \label{fig_9}
        \end{center}
\end{figure}

Considering the low sensitivity of the distance of disagreement (61 observed species) to the elemental sulphur abundance and because our improvement of the network concerns mainly the sulphur chemistry, we also assess the efficiency of our enhanced network by checking its agreement with only the observations of the S-bearing species detected in TMC-1 (CP). Figure \ref{fig_9} is the same as Figure \ref{fig_8} but considering only the 11 detected S-bearing gas-phase species in TMC-1 (CP). Those species are OCS, NS, HCS$^+$, CS, H$_2$CS, CCS, C$_3$S, SO, SO$_2$ \citep[see][for a review]{Agundez13}, HNCS and HSCN \citep{Adande10}. As expected, the distances of disagreement are much more sensitive to the elemental sulphur abundance than in the previous case, with differences going up to an average order of magnitude of 1.6 at some time steps. Moreover, the figure shows that the distance of disagreement for Model 1 and 2 never goes down to 0.7 whereas the other two present their minima under 0.4. Models 1 and 2 are therefore the 'worst' models which means the depleted elemental sulphur abundance of $8\times10^{-8}$ and $5\times10^{-7}$ do not allow the model to reproduce the observations of S-bearing species with our enhanced network.\\

However, at the times of their respective minimums, Models 3 and 4 reproduce all the S-bearing molecules considered. Model 4 is the one that presents the distance of disagreement with the lowest minimum of $D_4(t_{BF})=0.371$ with $t_{BF}=1.3\times10^6$ years. Furthermore, Model 3 allows us to obtain a minimum of the distance of disagreement that is nearly equal to the one found for Model 4, namely $D_3(t_{BF})=0.376$ with $t_{BF}=10^6$ years. In both cases, the time of 'best fit' is in the acceptable range for the age of TMC-1 (CP). These results suggest that our model is able to reproduce the observations of S-bearing species in TMC-1 (CP) using the cosmic elemental abundance of sulphur $X(S)_{ini}=1.5\times10^{-5}$ or an abundance 3 times lower.

\subsubsection{The case of HNCS and HSCN}

\begin{figure}
        \begin{center}
                \includegraphics[scale=0.14]{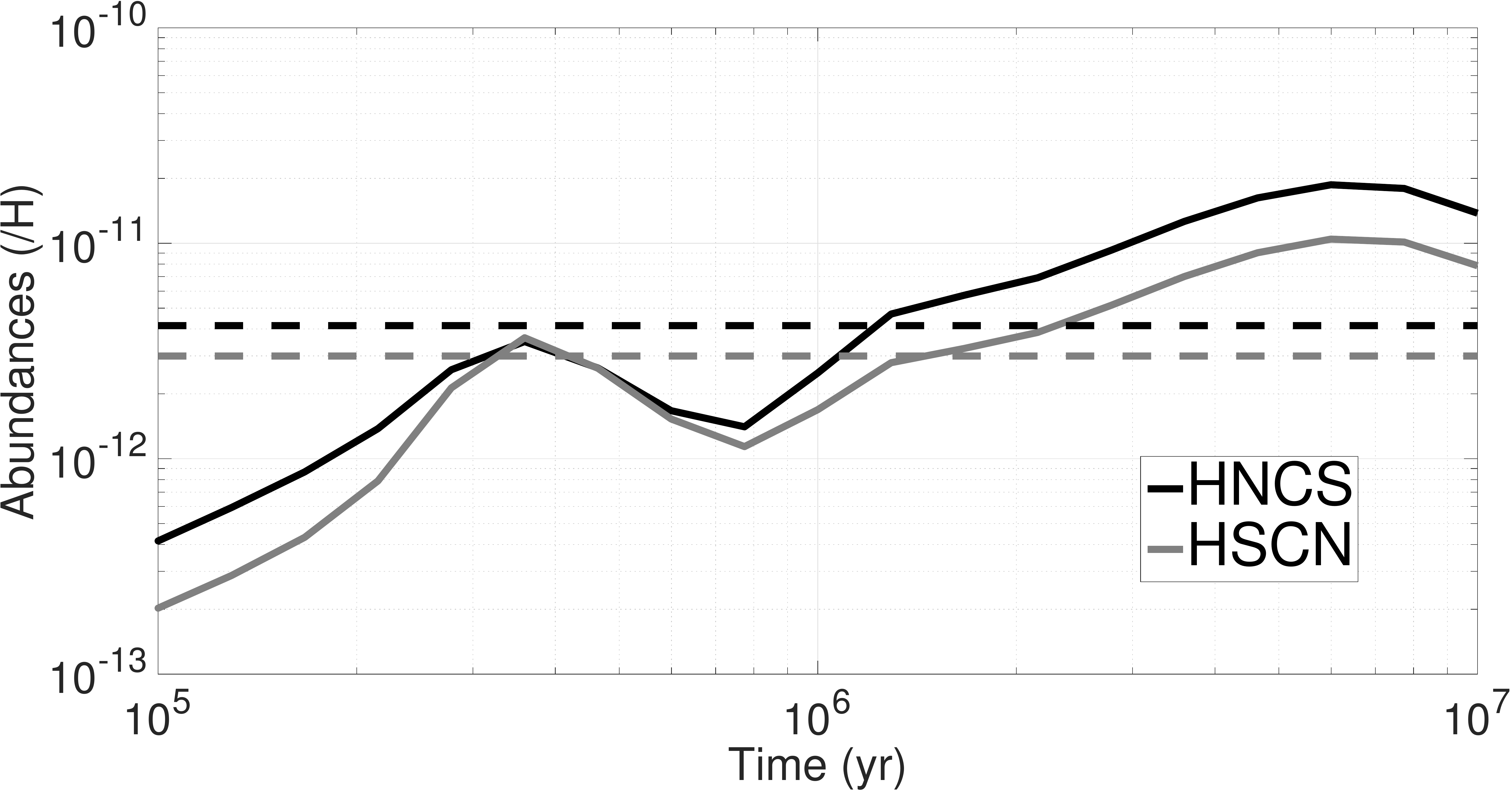}
                \caption{Comparison of the simulated gas-phase abundances of HNCS and HSCN with the observed ones in TMC-1 (CP) (horizontal dashed lines), relative to H.}
                \label{fig_10}
        \end{center}
\end{figure}

As our enhanced network was mainly built around HNCS and its isomers, it is important to check that our results fit well the observations of these species. \cite{Adande10} identified the $7_{0,7} \to 6_{0,6}$ and $8_{0,8} \to 7_{0,7}$ transitions of HNCS and HSCN towards TMC-1 (CP) and were able to derive a corresponding ratio $\frac{[\text{HNCS}]}{[\text{HSCN}]} = 1.4\pm0.7$ and abundances for HNCS and HSCN of respectively $4.15\times10^{-12}$ and $3\times10^{-12}$. Figure \ref{fig_10} shows the simulated abundances of HNCS and HSCN in the case of the cosmic elemental abundance for sulphur (Model 4) as well as their observed abundances in TMC-1 (CP) (horizontal dashed lines). At $t_{BF}=1.3\times10^6$ years, we obtain a ratio of 1.68 which seems to fit well with the observed value. Moreover, the figure shows that at this time, the differences between the observed and calculated abundances of HNCS and HSCN are less than a factor 1.2 and 1.1, respectively, which confirms the efficiency of the network to model the chemistry of HNCS. It should be noted that Model 3 also allows to reproduce well the observed abundance of HNCS and HSCN at $t_{BF}=10^6$ years.\\

\subsubsection{Comparison with observed S-bearing species on grains towards W33A}

\begin{figure}
        \begin{center}
                \includegraphics[scale=0.14]{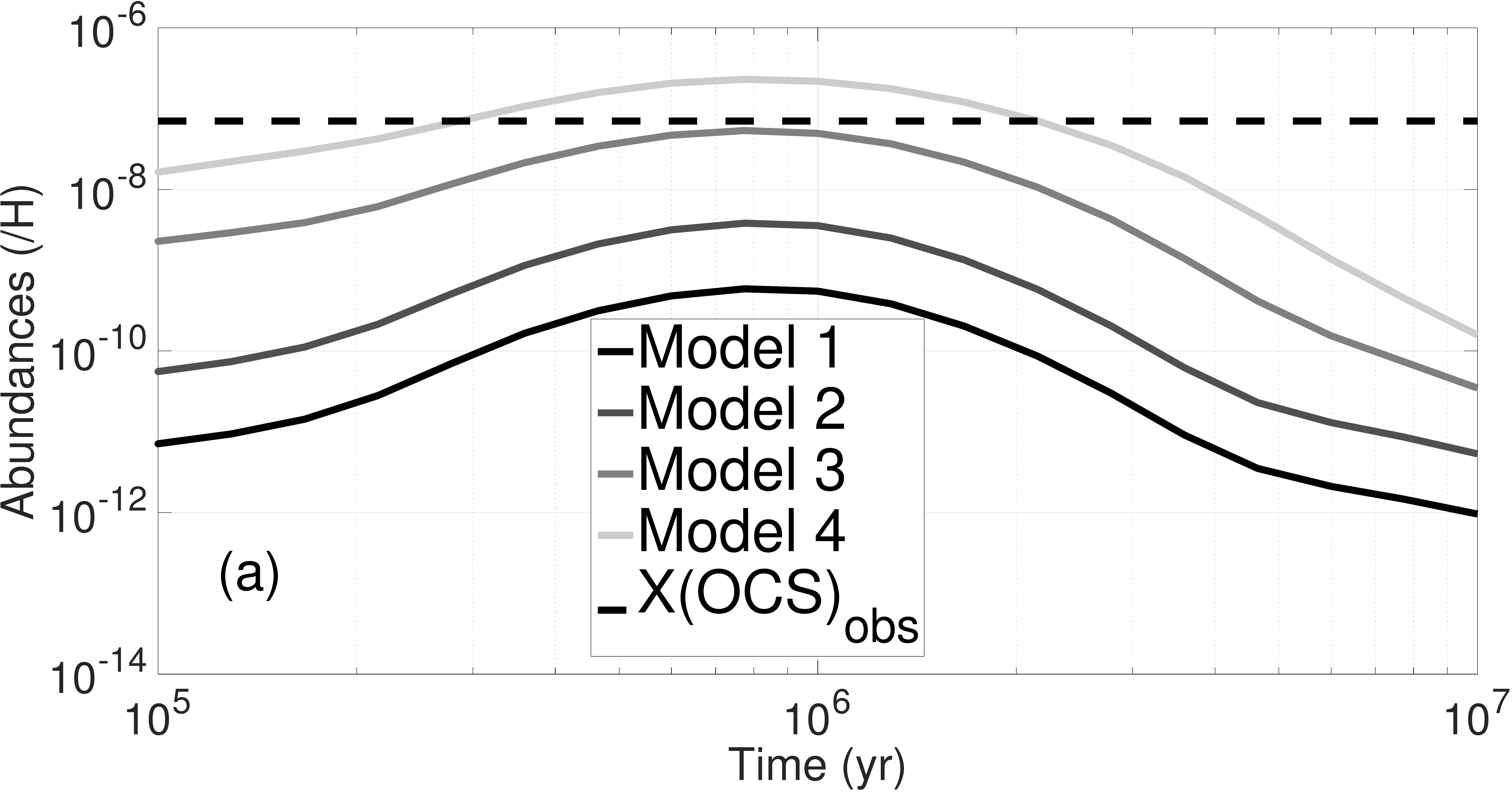}
                \includegraphics[scale=0.14]{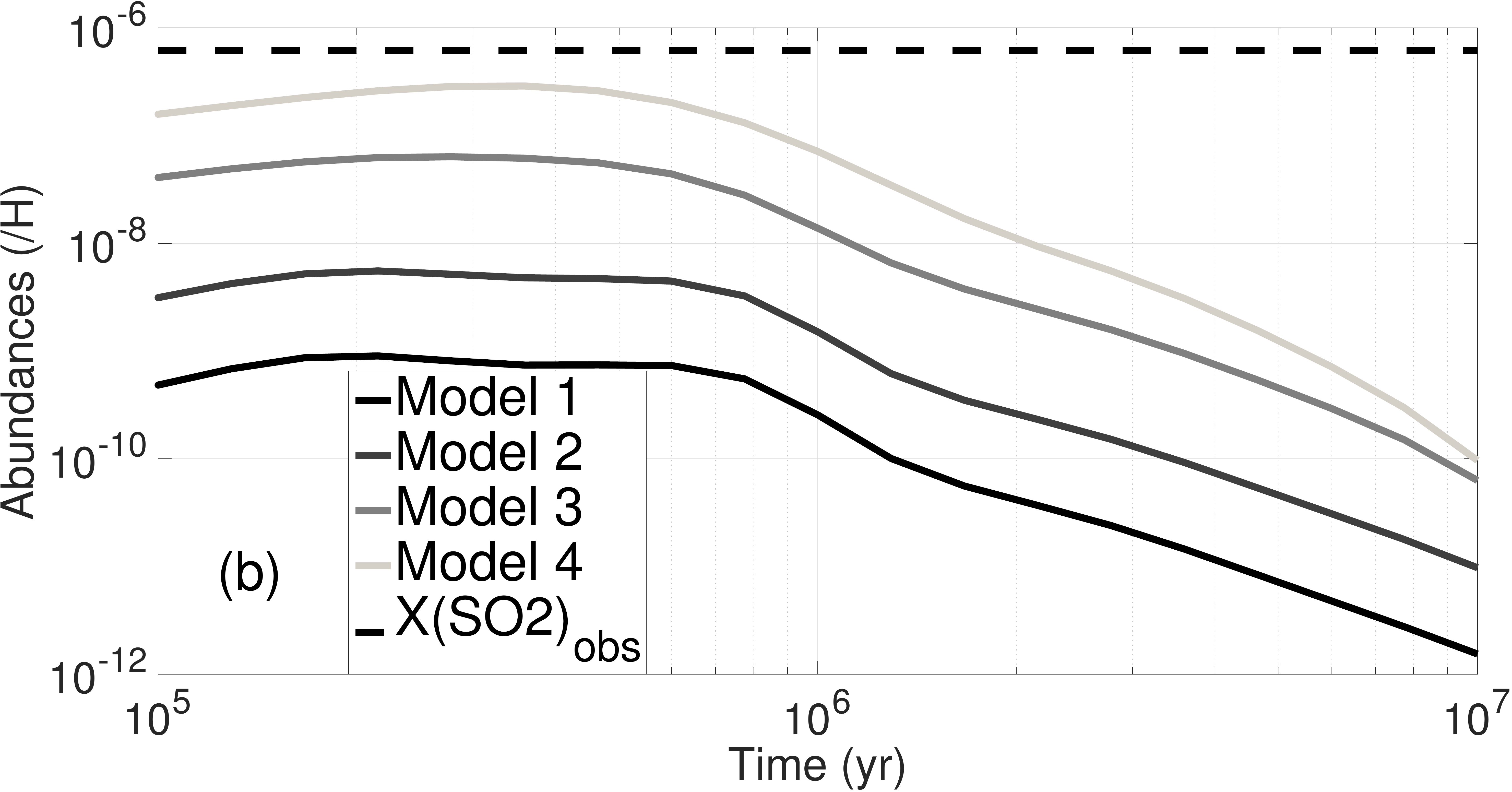}
                \caption{Comparison of the calculated abundances in icy grain bulks of OCS (a) and SO$_2$ (b) for Models 1, 2, 3 and 4 (solid lines) with the observed one towards W33A (dashed lines).}
                \label{fig_11}
        \end{center}
\end{figure}

Only two solid S-bearing species have been detected in icy grain bulks : OCS \citep{Palumbo97} and SO$_2$ \citep{Boogert97}. Both species have been detected towards the deeply embedded protostar W33A and we can therefore use the results of our Models in the dark cloud configuration to compare with these observations. In order to do so, we have to compare for a given species its observed abundance ($7\times10^{-8}$ and $6.2\times10^{-7}$ for OCS and SO$_2$, respectively) with the sum of its simulated abundances on the grain surface and in the grain bulk. For both species, Figure \ref{fig_11} displays the comparison between their simulated total abundances for our four Models in icy grain bulks and their observed ones between $10^5$ and $10^7$ years. The age of the giant molecular cloud W33 is indeed supposed to lie in that time frame \citep{Messineo15}. 

Figure \ref{fig_11} (a) shows that both Models 3 and 4 reproduce the observed abundance of OCS in icy grain bulks, at $7.74\times10^5$ and $2.15\times10^6$ years respectively. Moreover, considering the fact that most of the young stellar objects detected in W33 have an estimated age of a few million years \citep[see for example][]{Messineo15}, the most favored model is Model 4. Regarding SO$_2$, Figure \ref{fig_11} (b) shows that the model for which its modeled abundance in icy grain bulks comes the closest to its observed one is Model 4. Both comparisons suggest once again that our enhanced network needs an elemental sulphur abundance close to the cosmic one in order to best reproduce the observations. Note that we assume here that the detections of OCS and SO$_2$ ices are real although the two species are not in the list of "firmly detected" species of \citet{Boogert15}. The JWST mission may give us more constraints on this matter.

\subsection{Impacts of the use of the cosmic abundance of sulphur on the reservoirs}

In section \ref{chem_study} we studied the chemistry of the main S-bearing species using the elemental abundances listed in Table \ref{tab_1}, more particularly the depleted abundance of sulphur of $X(S)_{ini}=8\times10^{-8}$. Now that we determined that in order to reproduced S-bearing species observations in dark clouds our model needs an elemental abundance of sulphur close to the cosmic one, we need to assess the impact of the use of such an abundance on the results.\\

\begin{figure}
        \begin{center}
                \includegraphics[scale=0.14]{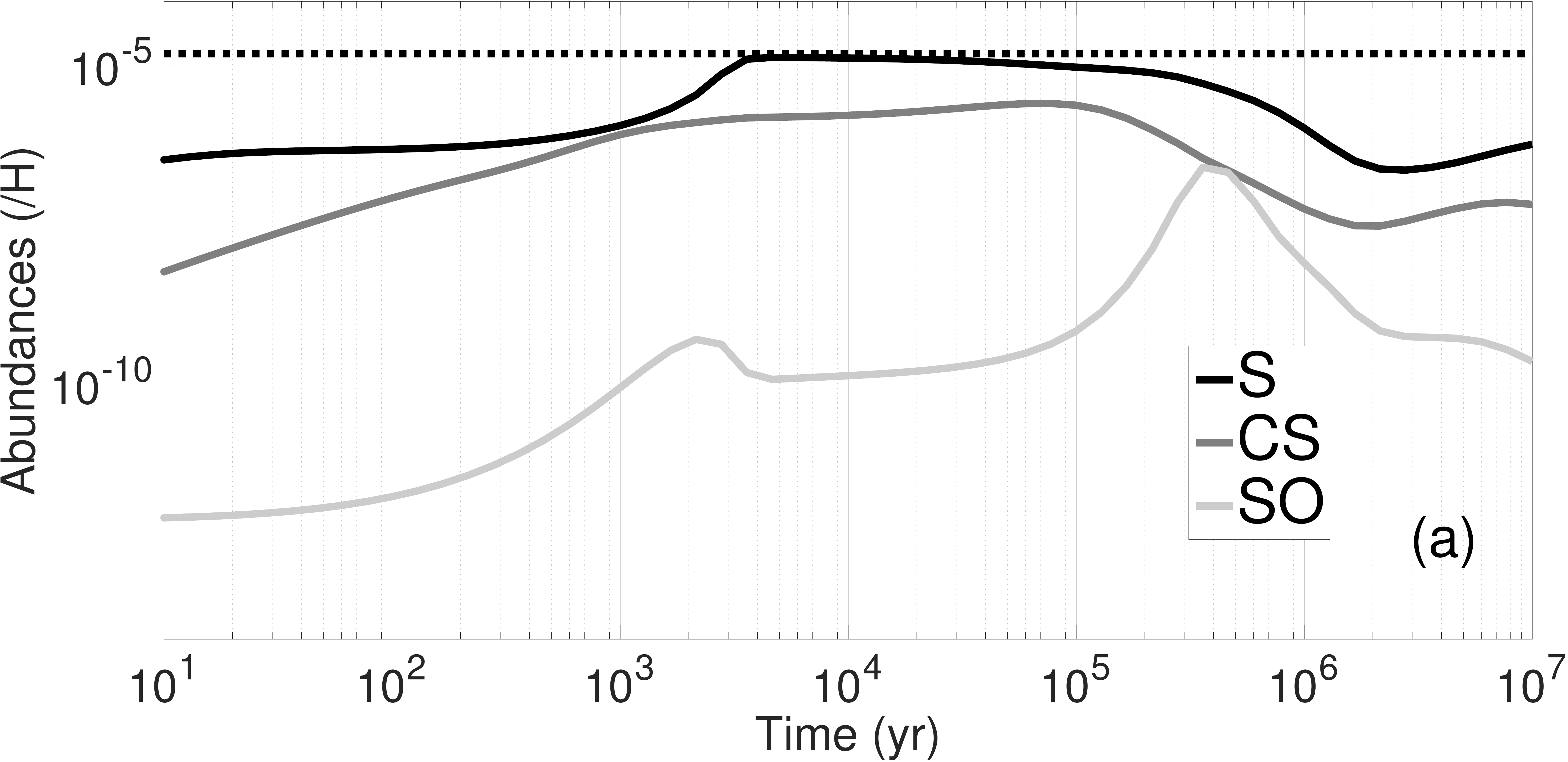}
                \includegraphics[scale=0.14]{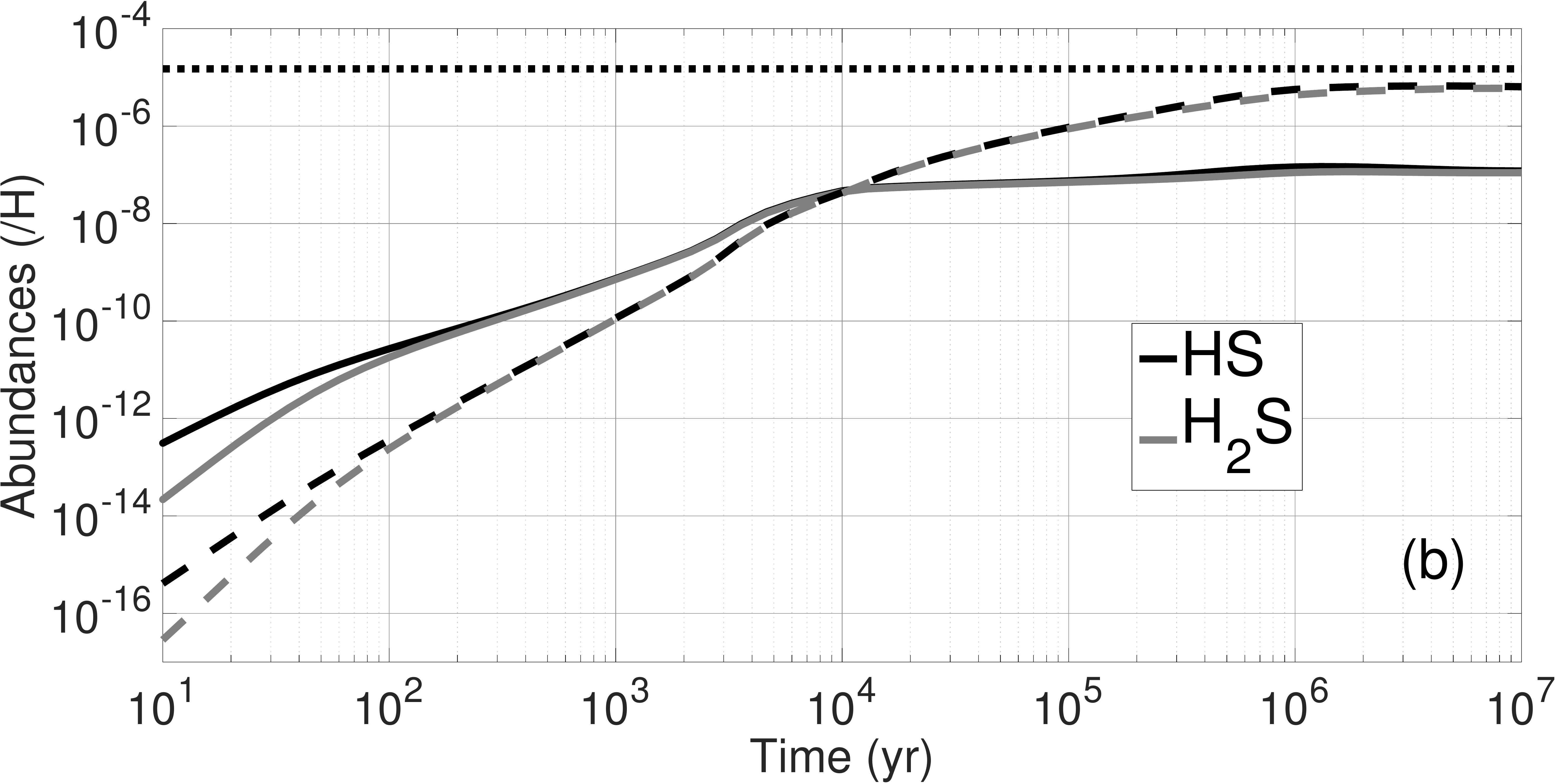}
                \caption{Abundances of main sulphur-bearing species relative to H as a function of time for dark cloud physical conditions : (a) in the gas phase, (b) on the grain surface (solid line) and bulk (dashed line). The dotted line represents the elemental abundance of sulphur (here, $X_{ini} = 1.5\times10^{-5}$).}
                \label{fig_111}
        \end{center}
\end{figure}

Figure \ref{fig_111} is the same as Figure \ref{fig_1} but for an elemental abundance of sulphur of $X_{ini} = 1.5\times10^{-5}$. What appears is that the differences are mainly quantitative and that the chemistry stay the same as described in section \ref{chem_study}, with the exception of the two following points: 

\begin{itemize}
	\item Adsorption of atomic S on the grain is much more efficient with $X_{ini} = 1.5\times10^{-5}$, which cause its abundance to drop much sooner than for $X(S)_{ini}=8\times10^{-8}$, near $3\times10^3$ years. However, atomic S still stays the main reservoir of sulphur between $2.8\times10^3$ and $4.6\times10^5$, containing at its maximum 88\% of the initial sulphur. \\
	\item With $X_{ini} = 1.5\times10^{-5}$, we can no longer consider SO as a main sulphur-bearing species as defined in section \ref{chem_study}. Indeed, at $3.6\times10^5$ years, time when its abundance peaks, it now only contains less than 2\% of the initial sulphur.
\end{itemize}

Hence, our results regarding the reservoirs of sulphur in dark clouds stay the same (with the exception of SO).

\section{Discussions and conclusions}

\subsection{About the elemental abundance of sulphur}

The comparison of our enhanced model with the gas-phase observations in TMC-1 (CP) (see section \ref{compobs}) favors the use of an elemental abundance of sulphur between $X(S)_{ini}=5\times10^{-6}$ and $X(S)_{ini}=1.5\times10^{-5}$. As most of the estimations of the cosmic abundance of sulphur lie between thess two values \citep{Federman93,Sofia94,Tieftrunk94,Ruffle99,Shalabiea01}, our model does not require additional depletion of sulphur. Moreover, our results on the comparison with the observations of OCS and SO$_2$ in icy grain bulks towards W33A confirm that a cosmic abundance of sulphur as the elemental sulphur abundance in our model is required to reproduce the observed abundances of these two species.\\

\subsection{About the reservoirs of sulphur}

Two main hypotheses exist to explain the observed depletion of sulphur in dark clouds: either the sulphur is in an as yet undiscovered form in the gas phase, which could be atomic sulphur, or it could be locked in icy grain bulks. The latter hypothesis is supported by the observation of H$_2$S as the most important S-bearing species in cometary ices \citep{Morvan00} which are thought to present chemical similarities with the ices processed during hot cores formation \citep{Irvine00}. Moreover, as hydrogenation is the most effective process in grains chemistry, models often predict that when sulphur atoms stick on the grain, they will consequently form H$_2$S \citep{Garrod07}. However, H$_2$S has never been detected in dark cloud ices. Additionally, upper limits to its column density towards three high mass protostars and three late type stars lying behind TMC-1 (CP) have been derived from observations by \cite{Smith91}. These limits, despite large uncertainties, are believed to be too small to account for the missing sulphur in dark clouds and alternative hypotheses have recently been proposed \citep{Jimenez11,Druard12,Martin16}.\\
 
As seen in section \ref{reservoirs}, the results of simulations with our enhanced sulphur network seem to support both of the main hypotheses about the reservoirs of sulphur, depending on the age of the observed dark cloud. Indeed, according to our results, if the age of the cloud is less than $5\times10^5$ years then atomic sulphur in the gas phase is the main reservoir of sulphur. However if the age of the cloud is greater, then our results show that because of hydrogenation of sulphur on the grain surface (the cycle between HS and H$_2$S enabled by the reaction-diffusion competition and the accumulation of these species in the bulk), the reservoirs of sulphur are nearly equally HS and H$_2$S in the grain bulk. The upper limit to the abundance of H$_2$S in ices provided by the observations is between $3\times10^{-7}$ and $3\times10^{-6}$ \citep[assuming an abundance of H$_2$O in ices of $10^{-4}$ and based on][]{Smith91}. It should be noted that these upper limits are derived using the infrared transition strengths obtained by \citet{Ferraro80} which has been obtained in pure H$_2$S ice. The intensity strengths can be notably different in interstellar ice analogues. As the total abundances of H$_2$S on the grains in Model 3 and 4 at the time of 'best fit' $10^6$ years are respectively $1.5\times10^{-6}$ and $4.4\times10^{-6}$, it appears that our enhanced network results are in agreement with the upper limits of \cite{Smith91} and therefore in favor of a joint HS and H$_2$S sulphur reservoir in icy grain bulks in dark clouds. This results is in agreement with the results depicted in appendix E of \citet{Furuya15}. Indeed, they also find that HS and H$_2$S are likely to be the main reservoirs of sulphur, notably due to the same hydrogenation loop described in equation (\ref{eq_2}). Constraints on the icy grain bulk abundance of HS would be useful to validate this hypothesis. However, The $v=0 \to v=1$ transition of HS lies around 2599 cm$^{-1}$ in the gaz phase \citep{Bernath83} and has never been observed in ice. The calculated IR strength (at DFT or MP2 level using Gaussian) is 3 orders of magnitude less efficient than water and 4 order of magnitude less efficient than OCS. Subsequently, even if solid HS is the reservoir of sulphur in dark clouds, the lack of sensibility of IR detection may prevent its observation.

\subsection{About the observability of HCS}

In order to assess the observability of HCS in the gas phase of TMC-1 (CP) (see section \ref{HCS}), we proceed using a local thermodynamic equilibrium calculation of the HCS emission spectrum in the 3 and 2 mm atmospheric windows with $T_{ex}=10$ K, a column density of HCS of $1\times10^{12}$ cm$^{-2}$ (corresponding to a gas phase abundance of HCS of approximately $10^{-10}$ as obtained with Model 4 in the time frame between 10$^5$ and 10$^7$ years), assuming a wide source (no beam dilution) and a linewidth of 1 km/s. We estimate that the brightest line would be the one at 80553.516 MHz with a main beam temperature of 2 mK. This result infers that for a $5\sigma$ detection of HCS in the gas phase, it would take more than a thousand hours of observations with the IRAM 30m. We therefore conclude that HCS is unlikely to be detected in the gas phase in dark clouds even if its abundance grows notably above $10^{-12}$ in our simulations. 

\section*{Acknowledgements}

This work has been founded by the European Research Council (Starting Grant 3DICE, grant agreement 336474). The authors are also grateful to the CNRS program "Physique et Chimie du Milieu Interstellaire" (PCMI) for partial funding of their work.




\bibliographystyle{mnras}
\bibliography{bibliography} 

\begin{thebibliography}{}
\makeatletter
\relax
\def\mn@urlcharsother{\let\do\@makeother \do\$\do\&\do\#\do\^\do\_\do\%\do\~}
\def\mn@doi{\begingroup\mn@urlcharsother \@ifnextchar [ {\mn@doi@}
  {\mn@doi@[]}}
\def\mn@doi@[#1]#2{\def\@tempa{#1}\ifx\@tempa\@empty \href
  {http://dx.doi.org/#2} {doi:#2}\else \href {http://dx.doi.org/#2} {#1}\fi
  \endgroup}
\def\mn@eprint#1#2{\mn@eprint@#1:#2::\@nil}
\def\mn@eprint@arXiv#1{\href {http://arxiv.org/abs/#1} {{\tt arXiv:#1}}}
\def\mn@eprint@dblp#1{\href {http://dblp.uni-trier.de/rec/bibtex/#1.xml}
  {dblp:#1}}
\def\mn@eprint@#1:#2:#3:#4\@nil{\def\@tempa {#1}\def\@tempb {#2}\def\@tempc
  {#3}\ifx \@tempc \@empty \let \@tempc \@tempb \let \@tempb \@tempa \fi \ifx
  \@tempb \@empty \def\@tempb {arXiv}\fi \@ifundefined
  {mn@eprint@\@tempb}{\@tempb:\@tempc}{\expandafter \expandafter \csname
  mn@eprint@\@tempb\endcsname \expandafter{\@tempc}}}

\bibitem[\protect\citeauthoryear{{Adande}, {Halfen}, {Ziurys}, {Quan}  \&
  {Herbst}}{{Adande} et~al.}{2010}]{Adande10}
{Adande} G.~R.,  {Halfen} D.~T.,  {Ziurys} L.~M.,  {Quan} D.,   {Herbst} E.,
  2010, \mn@doi [\apj] {10.1088/0004-637X/725/1/561}, \href
  {http://adsabs.harvard.edu/abs/2010ApJ...725..561A} {725, 561}

\bibitem[\protect\citeauthoryear{{Ag{\'u}ndez} \& {Wakelam}}{{Ag{\'u}ndez} \&
  {Wakelam}}{2013}]{Agundez13}
{Ag{\'u}ndez} M.,  {Wakelam} V.,  2013, \mn@doi [Chemical Reviews]
  {10.1021/cr4001176}, \href
  {http://adsabs.harvard.edu/abs/2013ChRv..113.8710A} {113, 8710}

\bibitem[\protect\citeauthoryear{Anicich}{Anicich}{2003}]{Anicich03}
Anicich V.~G.,  2003, JPL Publication 2003, 03-19 NASA.

\bibitem[\protect\citeauthoryear{Baulch et~al.,}{Baulch
  et~al.}{2005}]{Baulch05}
Baulch D.,  et~al., 2005, Journal of physical and chemical reference data, 34,
  757

\bibitem[\protect\citeauthoryear{Benson}{Benson}{1976}]{Benson76}
Benson S.,  1976, Second eddition, John Wiley \& Sons, New York

\bibitem[\protect\citeauthoryear{Bernath, Amano  \& Wong}{Bernath
  et~al.}{1983}]{Bernath83}
Bernath P.,  Amano T.,   Wong M.,  1983, Journal of Molecular Spectroscopy, 98,
  20

\bibitem[\protect\citeauthoryear{{Bockel{\'e}e-Morvan}
  et~al.,}{{Bockel{\'e}e-Morvan} et~al.}{2000}]{Morvan00}
{Bockel{\'e}e-Morvan} D.,  et~al., 2000, \aap, \href
  {http://adsabs.harvard.edu/abs/2000A%26A...353.1101B} {353, 1101}

\bibitem[\protect\citeauthoryear{{Boogert}, {Schutte}, {Helmich}, {Tielens}  \&
  {Wooden}}{{Boogert} et~al.}{1997}]{Boogert97}
{Boogert} A.~C.~A.,  {Schutte} W.~A.,  {Helmich} F.~P.,  {Tielens} A.~G.~G.~M.,
    {Wooden} D.~H.,  1997, \aap, \href
  {http://adsabs.harvard.edu/abs/1997A%26A...317..929B} {317, 929}

\bibitem[\protect\citeauthoryear{{Boogert}, {Gerakines}  \&
  {Whittet}}{{Boogert} et~al.}{2015}]{Boogert15}
{Boogert} A.~C.~A.,  {Gerakines} P.~A.,   {Whittet} D.~C.~B.,  2015, \mn@doi
  [\araa] {10.1146/annurev-astro-082214-122348}, \href
  {http://adsabs.harvard.edu/abs/2015ARA%26A..53..541B} {53, 541}

\bibitem[\protect\citeauthoryear{{Cernicharo} et~al.,}{{Cernicharo}
  et~al.}{2011}]{Cernicharo11}
{Cernicharo} J.,  et~al., 2011, \mn@doi [\aap] {10.1051/0004-6361/201016216},
  \href {http://adsabs.harvard.edu/abs/2011A%26A...531A.103C} {531, A103}

\bibitem[\protect\citeauthoryear{Chabot et~al.,}{Chabot
  et~al.}{2010}]{Chabot10}
Chabot M.,  et~al., 2010, Astronomy \& Astrophysics, 524, A39

\bibitem[\protect\citeauthoryear{{Chang}, {Cuppen}  \& {Herbst}}{{Chang}
  et~al.}{2007}]{Chang07}
{Chang} Q.,  {Cuppen} H.~M.,   {Herbst} E.,  2007, \mn@doi [\aap]
  {10.1051/0004-6361:20077423}, \href
  {http://adsabs.harvard.edu/abs/2007A%26A...469..973C} {469, 973}

\bibitem[\protect\citeauthoryear{{Charnley}}{{Charnley}}{1997}]{Charnley97}
{Charnley} S.~B.,  1997, \apj, \href
  {http://adsabs.harvard.edu/abs/1997ApJ...481..396C} {481, 396}

\bibitem[\protect\citeauthoryear{{Druard} \& {Wakelam}}{{Druard} \&
  {Wakelam}}{2012}]{Druard12}
{Druard} C.,  {Wakelam} V.,  2012, \mn@doi [\mnras]
  {10.1111/j.1365-2966.2012.21712.x}, \href
  {http://adsabs.harvard.edu/abs/2012MNRAS.426..354D} {426, 354}

\bibitem[\protect\citeauthoryear{{Federman}, {Sheffer}, {Lambert}  \&
  {Gilliland}}{{Federman} et~al.}{1993}]{Federman93}
{Federman} S.~R.,  {Sheffer} Y.,  {Lambert} D.~L.,   {Gilliland} R.~L.,  1993,
  \mn@doi [\apjl] {10.1086/186957}, \href
  {http://adsabs.harvard.edu/abs/1993ApJ...413L..51F} {413, L51}

\bibitem[\protect\citeauthoryear{Ferraro, Sill  \& Fink}{Ferraro
  et~al.}{1980}]{Ferraro80}
Ferraro J.~R.,  Sill G.,   Fink U.,  1980, Applied Spectroscopy, 34, 525

\bibitem[\protect\citeauthoryear{Florescu-Mitchell \&
  Mitchell}{Florescu-Mitchell \& Mitchell}{2006}]{Florescu06}
Florescu-Mitchell A.~I.,  Mitchell J. B.~A.,  2006, Phys. Rep., 430, 277

\bibitem[\protect\citeauthoryear{Fournier, Shuman, Melko, Ard  \&
  Viggiano}{Fournier et~al.}{2013}]{Fournier13}
Fournier J.~A.,  Shuman N.~S.,  Melko J.~J.,  Ard S.~G.,   Viggiano A.~A.,
  2013, J. Chem. Phys., 138, 154201

\bibitem[\protect\citeauthoryear{{Frerking}, {Linke}  \& {Thaddeus}}{{Frerking}
  et~al.}{1979}]{Frerking79}
{Frerking} M.~A.,  {Linke} R.~A.,   {Thaddeus} P.,  1979, \mn@doi [\apjl]
  {10.1086/183126}, \href {http://adsabs.harvard.edu/abs/1979ApJ...234L.143F}
  {234, L143}

\bibitem[\protect\citeauthoryear{{Furuya}, {Aikawa}, {Hincelin}, {Hassel},
  {Bergin}, {Vasyunin}  \& {Herbst}}{{Furuya} et~al.}{2015}]{Furuya15}
{Furuya} K.,  {Aikawa} Y.,  {Hincelin} U.,  {Hassel} G.~E.,  {Bergin} E.~A.,
  {Vasyunin} A.~I.,   {Herbst} E.,  2015, \mn@doi [\aap]
  {10.1051/0004-6361/201527050}, \href
  {http://adsabs.harvard.edu/abs/2015A%26A...584A.124F} {584, A124}

\bibitem[\protect\citeauthoryear{Galland, Caralp, Rayez, Hannachi, Loison,
  Dorthe  \& Bergeat}{Galland et~al.}{2001}]{Galland01}
Galland N.,  Caralp F.,  Rayez M.-T.,  Hannachi Y.,  Loison J.-C.,  Dorthe G.,
   Bergeat A.,  2001, The Journal of Physical Chemistry A, 105, 9893

\bibitem[\protect\citeauthoryear{Galland, Caralp, Hannachi, Bergeat  \&
  Loison}{Galland et~al.}{2003}]{Galland03}
Galland N.,  Caralp F.,  Hannachi Y.,  Bergeat A.,   Loison J.-C.,  2003, The
  Journal of Physical Chemistry A, 107, 5419

\bibitem[\protect\citeauthoryear{{Garozzo}, {Fulvio}, {Kanuchova}, {Palumbo}
  \& {Strazzulla}}{{Garozzo} et~al.}{2010}]{Garozzo10}
{Garozzo} M.,  {Fulvio} D.,  {Kanuchova} Z.,  {Palumbo} M.~E.,   {Strazzulla}
  G.,  2010, \mn@doi [\aap] {10.1051/0004-6361/200913040}, \href
  {http://adsabs.harvard.edu/abs/2010A%26A...509A..67G} {509, A67}

\bibitem[\protect\citeauthoryear{Garrod \& Herbst}{Garrod \&
  Herbst}{2006}]{Garrod06}
Garrod R.,  Herbst E.,  2006, Astronomy \& Astrophysics, 457, 927

\bibitem[\protect\citeauthoryear{{Garrod} \& {Pauly}}{{Garrod} \&
  {Pauly}}{2011}]{Garrod11}
{Garrod} R.~T.,  {Pauly} T.,  2011, \mn@doi [\apj]
  {10.1088/0004-637X/735/1/15}, \href
  {http://adsabs.harvard.edu/abs/2011ApJ...735...15G} {735, 15}

\bibitem[\protect\citeauthoryear{{Garrod}, {Wakelam}  \& {Herbst}}{{Garrod}
  et~al.}{2007}]{Garrod07}
{Garrod} R.~T.,  {Wakelam} V.,   {Herbst} E.,  2007, \mn@doi [\aap]
  {10.1051/0004-6361:20066704}, \href
  {http://adsabs.harvard.edu/abs/2007A%26A...467.1103G} {467, 1103}

\bibitem[\protect\citeauthoryear{Georgievskii \& Klippenstein}{Georgievskii \&
  Klippenstein}{2005}]{Georgievskii05}
Georgievskii Y.,  Klippenstein S.~J.,  2005, The Journal of chemical physics,
  122, 194103

\bibitem[\protect\citeauthoryear{Georgievskii \& Klippenstein}{Georgievskii \&
  Klippenstein}{2007}]{Georgievskii07}
Georgievskii Y.,  Klippenstein S.~J.,  2007, The Journal of Physical Chemistry
  A, 111, 3802

\bibitem[\protect\citeauthoryear{{Gibb}, {Nummelin}, {Irvine}, {Whittet}  \&
  {Bergman}}{{Gibb} et~al.}{2000}]{Gibb00}
{Gibb} E.,  {Nummelin} A.,  {Irvine} W.~M.,  {Whittet} D.~C.~B.,   {Bergman}
  P.,  2000, \mn@doi [\apj] {10.1086/317805}, \href
  {http://adsabs.harvard.edu/abs/2000ApJ...545..309G} {545, 309}

\bibitem[\protect\citeauthoryear{Glowacki, Liang, Morley, Pilling  \&
  Robertson}{Glowacki et~al.}{2012}]{Glowacki12}
Glowacki D.~R.,  Liang C.-H.,  Morley C.,  Pilling M.~J.,   Robertson S.~H.,
  2012, The Journal of Physical Chemistry A, 116, 9545

\bibitem[\protect\citeauthoryear{Gonzalez, Hijazo, Novoa  \& Sayos}{Gonzalez
  et~al.}{1996}]{Gonzalez96}
Gonzalez M.,  Hijazo J.,  Novoa J.~J.,   Sayos R.,  1996, J. Chem. Phys., 105,
  10999

\bibitem[\protect\citeauthoryear{{Graedel}, {Langer}  \& {Frerking}}{{Graedel}
  et~al.}{1982}]{Graedel82}
{Graedel} T.~E.,  {Langer} W.~D.,   {Frerking} M.~A.,  1982, \mn@doi [\apjs]
  {10.1086/190780}, \href {http://adsabs.harvard.edu/abs/1982ApJS...48..321G}
  {48, 321}

\bibitem[\protect\citeauthoryear{{Gratier}, {Majumdar}, {Ohishi}, {Roueff},
  {Loison}, {Hickson}  \& {Wakelam}}{{Gratier} et~al.}{2016}]{Gratier16}
{Gratier} P.,  {Majumdar} L.,  {Ohishi} M.,  {Roueff} E.,  {Loison} J.~C.,
  {Hickson} K.~M.,   {Wakelam} V.,  2016, \mn@doi [\apjs]
  {10.3847/0067-0049/225/2/25}, \href
  {http://adsabs.harvard.edu/abs/2016ApJS..225...25G} {225, 25}

\bibitem[\protect\citeauthoryear{Gronowski \& Kolos}{Gronowski \&
  Kolos}{2014}]{Gronowski14}
Gronowski M.,  Kolos R.,  2014, ApJ, 792, 89

\bibitem[\protect\citeauthoryear{{Halfen}, {Ziurys}, {Br{\"u}nken}, {Gottlieb},
  {McCarthy}  \& {Thaddeus}}{{Halfen} et~al.}{2009}]{Halfen09}
{Halfen} D.~T.,  {Ziurys} L.~M.,  {Br{\"u}nken} S.,  {Gottlieb} C.~A.,
  {McCarthy} M.~C.,   {Thaddeus} P.,  2009, \mn@doi [\apjl]
  {10.1088/0004-637X/702/2/L124}, \href
  {http://adsabs.harvard.edu/abs/2009ApJ...702L.124H} {702, L124}

\bibitem[\protect\citeauthoryear{Hama \& Watanabe}{Hama \&
  Watanabe}{2013}]{Hama13}
Hama T.,  Watanabe N.,  2013, Chemical reviews, 113, 8783

\bibitem[\protect\citeauthoryear{Hamberg et~al.,}{Hamberg
  et~al.}{2007}]{Hamberg07}
Hamberg M.,  et~al., 2007, \mn@doi [Mol. Phys.] {10.1080/00268970701206642},
  105, 899

\bibitem[\protect\citeauthoryear{Hamberg et~al.,}{Hamberg
  et~al.}{2014}]{Hamberg14}
Hamberg M.,  et~al., 2014, \mn@doi [The Journal of Physical Chemistry A]
  {10.1021/jp5032306}, 118, 6034

\bibitem[\protect\citeauthoryear{{Hasegawa}, {Herbst}  \& {Leung}}{{Hasegawa}
  et~al.}{1992}]{Hasegawa92}
{Hasegawa} T.~I.,  {Herbst} E.,   {Leung} C.~M.,  1992, \mn@doi [\apjs]
  {10.1086/191713}, \href {http://adsabs.harvard.edu/abs/1992ApJS...82..167H}
  {82, 167}

\bibitem[\protect\citeauthoryear{{Hatchell}, {Thompson}, {Millar}  \&
  {MacDonald}}{{Hatchell} et~al.}{1998}]{Hatchell98}
{Hatchell} J.,  {Thompson} M.~A.,  {Millar} T.~J.,   {MacDonald} G.~H.,  1998,
  \aap, \href {http://adsabs.harvard.edu/abs/1998A%26A...338..713H} {338, 713}

\bibitem[\protect\citeauthoryear{Herbst, Terzieva  \& Talbi}{Herbst
  et~al.}{2000}]{Herbst00}
Herbst E.,  Terzieva R.,   Talbi D.,  2000, \mn@doi [MNRAS]
  {10.1046/j.1365-8711.2000.03103.x}, 311, 869

\bibitem[\protect\citeauthoryear{Hickson, Loison, Cavali{\'e}, H{\'e}brard  \&
  Dobrijevic}{Hickson et~al.}{2014}]{Hickson14}
Hickson K.,  Loison J.,  Cavali{\'e} T.,  H{\'e}brard E.,   Dobrijevic M.,
  2014, Astronomy \& Astrophysics, 572, A58

\bibitem[\protect\citeauthoryear{Hickson, Loison, Nunez-Reyes  \&
  M{\'e}reau}{Hickson et~al.}{2016a}]{Hickson16b}
Hickson K.~M.,  Loison J.-C.,  Nunez-Reyes D.,   M{\'e}reau R.,  2016a, The
  Journal of Physical Chemistry Letters

\bibitem[\protect\citeauthoryear{{Hickson}, {Wakelam}  \& {Loison}}{{Hickson}
  et~al.}{2016b}]{Hickson16}
{Hickson} K.~M.,  {Wakelam} V.,   {Loison} J.-C.,  2016b, \mn@doi [Molecular
  Astrophysics] {10.1016/j.molap.2016.03.001}, \href
  {http://adsabs.harvard.edu/abs/2016MolAs...3....1H} {3, 1}

\bibitem[\protect\citeauthoryear{{Hincelin}, {Wakelam}, {Hersant},
  {Guilloteau}, {Loison}, {Honvault}  \& {Troe}}{{Hincelin}
  et~al.}{2011}]{Hincelin11}
{Hincelin} U.,  {Wakelam} V.,  {Hersant} F.,  {Guilloteau} S.,  {Loison} J.~C.,
   {Honvault} P.,   {Troe} J.,  2011, \mn@doi [\aap]
  {10.1051/0004-6361/201016328}, \href
  {http://adsabs.harvard.edu/abs/2011A%26A...530A..61H} {530, A61}

\bibitem[\protect\citeauthoryear{Hippler \& Viskolcz}{Hippler \&
  Viskolcz}{2002}]{Hippler02}
Hippler H.,  Viskolcz B.,  2002, Physical Chemistry Chemical Physics, 4, 4663

\bibitem[\protect\citeauthoryear{{Holdship} et~al.,}{{Holdship}
  et~al.}{2016}]{Holdship16}
{Holdship} J.,  et~al., 2016, \mn@doi [\mnras] {10.1093/mnras/stw1977}, \href
  {http://adsabs.harvard.edu/abs/2016MNRAS.tmp.1090H} {}

\bibitem[\protect\citeauthoryear{{Irvine}, {Schloerb}, {Crovisier}, {Fegley}
  \& {Mumma}}{{Irvine} et~al.}{2000}]{Irvine00}
{Irvine} W.~M.,  {Schloerb} F.~P.,  {Crovisier} J.,  {Fegley} Jr. B.,   {Mumma}
  M.~J.,  2000, Protostars and Planets IV, \href
  {http://adsabs.harvard.edu/abs/2000prpl.conf.1159I} {p.~1159}

\bibitem[\protect\citeauthoryear{{Jenkins}}{{Jenkins}}{2009}]{Jenkins09}
{Jenkins} E.~B.,  2009, \mn@doi [\apj] {10.1088/0004-637X/700/2/1299}, \href
  {http://adsabs.harvard.edu/abs/2009ApJ...700.1299J} {700, 1299}

\bibitem[\protect\citeauthoryear{{Jim{\'e}nez-Escobar} \& {Mu{\~n}oz
  Caro}}{{Jim{\'e}nez-Escobar} \& {Mu{\~n}oz Caro}}{2011}]{Jimenez11}
{Jim{\'e}nez-Escobar} A.,  {Mu{\~n}oz Caro} G.~M.,  2011, \mn@doi [\aap]
  {10.1051/0004-6361/201014821}, \href
  {http://adsabs.harvard.edu/abs/2011A%26A...536A..91J} {536, A91}

\bibitem[\protect\citeauthoryear{{Karssemeijer} \& {Cuppen}}{{Karssemeijer} \&
  {Cuppen}}{2014}]{Karssemeijer14}
{Karssemeijer} L.~J.,  {Cuppen} H.~M.,  2014, \mn@doi [\aap]
  {10.1051/0004-6361/201424792}, \href
  {http://adsabs.harvard.edu/abs/2014A%26A...569A.107K} {569, A107}

\bibitem[\protect\citeauthoryear{{Kolesnikov{\'a}}, {Tercero}, {Cernicharo},
  {Alonso}, {Daly}, {Gordon}  \& {Shipman}}{{Kolesnikov{\'a}}
  et~al.}{2014}]{Kolesnikov14}
{Kolesnikov{\'a}} L.,  {Tercero} B.,  {Cernicharo} J.,  {Alonso} J.~L.,  {Daly}
  A.~M.,  {Gordon} B.~P.,   {Shipman} S.~T.,  2014, \mn@doi [\apjl]
  {10.1088/2041-8205/784/1/L7}, \href
  {http://adsabs.harvard.edu/abs/2014ApJ...784L...7K} {784, L7}

\bibitem[\protect\citeauthoryear{Korth \& Grimme}{Korth \&
  Grimme}{2009}]{Korth09}
Korth M.,  Grimme S.,  2009, Journal of chemical theory and computation, 5, 993

\bibitem[\protect\citeauthoryear{Kurylo, Peterson  \& Braun}{Kurylo
  et~al.}{1971}]{Kurylo71}
Kurylo M.~J.,  Peterson N.~C.,   Braun W.,  1971, The Journal of Chemical
  Physics, 54, 943

\bibitem[\protect\citeauthoryear{{Lada}, {Bally}  \& {Stark}}{{Lada}
  et~al.}{1991}]{Lada91}
{Lada} E.~A.,  {Bally} J.,   {Stark} A.~A.,  1991, \mn@doi [\apj]
  {10.1086/169708}, \href {http://adsabs.harvard.edu/abs/1991ApJ...368..432L}
  {368, 432}

\bibitem[\protect\citeauthoryear{Leonori, Occhiogrosso, Balucani, Bucci,
  Petrucci  \& Casavecchia}{Leonori et~al.}{2011}]{Leonori11}
Leonori F.,  Occhiogrosso A.,  Balucani N.,  Bucci A.,  Petrucci R.,
  Casavecchia P.,  2011, The Journal of Physical Chemistry Letters, 3, 75

\bibitem[\protect\citeauthoryear{Leonori, Balucani, Nevrly, Bergeat,
  Falcinelli, Vanuzzo, Casavecchia  \& Cavallotti}{Leonori
  et~al.}{2015}]{Leonori15}
Leonori F.,  Balucani N.,  Nevrly V.,  Bergeat A.,  Falcinelli S.,  Vanuzzo G.,
   Casavecchia P.,   Cavallotti C.,  2015, The Journal of Physical Chemistry C,
  119, 14632

\bibitem[\protect\citeauthoryear{Lilenfeld \& Richardson}{Lilenfeld \&
  Richardson}{1977}]{Lilenfeld77}
Lilenfeld H.~V.,  Richardson R.~J.,  1977, J. Chem. Phys., 67, 3991

\bibitem[\protect\citeauthoryear{{Linke}, {Frerking}  \& {Thaddeus}}{{Linke}
  et~al.}{1979}]{Linke79}
{Linke} R.~A.,  {Frerking} M.~A.,   {Thaddeus} P.,  1979, \mn@doi [\apjl]
  {10.1086/183125}, \href {http://adsabs.harvard.edu/abs/1979ApJ...234L.139L}
  {234, L139}

\bibitem[\protect\citeauthoryear{{Lique}, {Cernicharo}  \& {Cox}}{{Lique}
  et~al.}{2006}]{Lique06}
{Lique} F.,  {Cernicharo} J.,   {Cox} P.,  2006, \mn@doi [\apj]
  {10.1086/508978}, \href {http://adsabs.harvard.edu/abs/2006ApJ...653.1342L}
  {653, 1342}

\bibitem[\protect\citeauthoryear{Loison, Wakelam, Hickson, Bergeat  \&
  Mereau}{Loison et~al.}{2014}]{Loison14}
Loison J.-C.,  Wakelam V.,  Hickson K.~M.,  Bergeat A.,   Mereau R.,  2014,
  \mn@doi [MNRAS] {10.1093/mnras/stt1956}, 437, 930

\bibitem[\protect\citeauthoryear{{Loison} et~al.,}{{Loison}
  et~al.}{2016}]{Loison16}
{Loison} J.-C.,  et~al., 2016, \mn@doi [\mnras] {10.1093/mnras/stv2866}, \href
  {http://adsabs.harvard.edu/abs/2016MNRAS.456.4101L} {456, 4101}

\bibitem[\protect\citeauthoryear{{Majumdar}, {Gratier}, {Vidal}, {Wakelam},
  {Loison}, {Hickson}  \& {Caux}}{{Majumdar} et~al.}{2016}]{Majumdar16}
{Majumdar} L.,  {Gratier} P.,  {Vidal} T.,  {Wakelam} V.,  {Loison} J.-C.,
  {Hickson} K.~M.,   {Caux} E.,  2016, \mn@doi [\mnras] {10.1093/mnras/stw457},
  \href {http://adsabs.harvard.edu/abs/2016MNRAS.458.1859M} {458, 1859}

\bibitem[\protect\citeauthoryear{{Mart{\'{\i}}n-Dom{\'e}nech},
  {Jim{\'e}nez-Serra}, {Mu{\~n}oz Caro}, {M{\"u}ller}, {Occhiogrosso}, {Testi},
  {Woods}  \& {Viti}}{{Mart{\'{\i}}n-Dom{\'e}nech} et~al.}{2016}]{Martin16}
{Mart{\'{\i}}n-Dom{\'e}nech} R.,  {Jim{\'e}nez-Serra} I.,  {Mu{\~n}oz Caro}
  G.~M.,  {M{\"u}ller} H.~S.~P.,  {Occhiogrosso} A.,  {Testi} L.,  {Woods}
  P.~M.,   {Viti} S.,  2016, \mn@doi [\aap] {10.1051/0004-6361/201526271},
  \href {http://adsabs.harvard.edu/abs/2016A%26A...585A.112M} {585, A112}

\bibitem[\protect\citeauthoryear{{Matthews}, {MacLeod}, {Broten}, {Madden}  \&
  {Friberg}}{{Matthews} et~al.}{1987}]{Matthews87}
{Matthews} H.~E.,  {MacLeod} J.~M.,  {Broten} N.~W.,  {Madden} S.~C.,
  {Friberg} P.,  1987, \mn@doi [\apj] {10.1086/165166}, \href
  {http://adsabs.harvard.edu/abs/1987ApJ...315..646M} {315, 646}

\bibitem[\protect\citeauthoryear{{McGonagle}, {Irvine}  \&
  {Ohishi}}{{McGonagle} et~al.}{1994}]{McGonagle94}
{McGonagle} D.,  {Irvine} W.~M.,   {Ohishi} M.,  1994, \mn@doi [\apj]
  {10.1086/173755}, \href {http://adsabs.harvard.edu/abs/1994ApJ...422..621M}
  {422, 621}

\bibitem[\protect\citeauthoryear{Mendes et~al.,}{Mendes
  et~al.}{2012}]{Mendes12}
Mendes M.~B.,  et~al., 2012, The Astrophysical Journal Letters, 746, L8

\bibitem[\protect\citeauthoryear{{Messineo} et~al.,}{{Messineo}
  et~al.}{2015}]{Messineo15}
{Messineo} M.,  et~al., 2015, \mn@doi [\apj] {10.1088/0004-637X/805/2/110},
  \href {http://adsabs.harvard.edu/abs/2015ApJ...805..110M} {805, 110}

\bibitem[\protect\citeauthoryear{Montaigne et~al.,}{Montaigne
  et~al.}{2005}]{Montaigne05}
Montaigne H.,  et~al., 2005, Astrophys. J., 631, 653

\bibitem[\protect\citeauthoryear{{Neufeld}, {Wolfire}  \& {Schilke}}{{Neufeld}
  et~al.}{2005}]{Neufeld05}
{Neufeld} D.~A.,  {Wolfire} M.~G.,   {Schilke} P.,  2005, \mn@doi [\apj]
  {10.1086/430663}, \href {http://adsabs.harvard.edu/abs/2005ApJ...628..260N}
  {628, 260}

\bibitem[\protect\citeauthoryear{Nobes \& Radom}{Nobes \&
  Radom}{1981}]{Nobes81}
Nobes R.~H.,  Radom L.,  1981, \mn@doi [Chem. Phys.]
  {http://dx.doi.org/10.1016/0301-0104(81)80102-4}, 60, 1

\bibitem[\protect\citeauthoryear{Oehlers, Wagner, Ziemer, Temps  \&
  Dobe}{Oehlers et~al.}{2000}]{Oehlers00}
Oehlers C.,  Wagner H.~G.,  Ziemer H.,  Temps F.,   Dobe S.,  2000, The Journal
  of Physical Chemistry A, 104, 10500

\bibitem[\protect\citeauthoryear{{Palumbo}, {Geballe}  \& {Tielens}}{{Palumbo}
  et~al.}{1997}]{Palumbo97}
{Palumbo} M.~E.,  {Geballe} T.~R.,   {Tielens} A.~G.~G.~M.,  1997, \apj, \href
  {http://adsabs.harvard.edu/abs/1997ApJ...479..839P} {479, 839}

\bibitem[\protect\citeauthoryear{Peng, Hu  \& Marshall}{Peng
  et~al.}{1999}]{Peng99}
Peng J.,  Hu X.,   Marshall P.,  1999, The Journal of Physical Chemistry A,
  103, 5307

\bibitem[\protect\citeauthoryear{{Penzias}, {Solomon}, {Wilson}  \&
  {Jefferts}}{{Penzias} et~al.}{1971}]{Penzias71}
{Penzias} A.~A.,  {Solomon} P.~M.,  {Wilson} R.~W.,   {Jefferts} K.~B.,  1971,
  \mn@doi [\apjl] {10.1086/180784}, \href
  {http://adsabs.harvard.edu/abs/1971ApJ...168L..53P} {168, L53}

\bibitem[\protect\citeauthoryear{Peters, Duflot, Wiesenfeld  \& Toubin}{Peters
  et~al.}{2013}]{Peters13}
Peters P.~S.,  Duflot D.,  Wiesenfeld L.,   Toubin C.,  2013, The Journal of
  chemical physics, 139, 164310

\bibitem[\protect\citeauthoryear{Plessis, Carrasco, Dobrijevic  \&
  Pernot}{Plessis et~al.}{2012}]{Plessis12}
Plessis S.,  Carrasco N.,  Dobrijevic M.,   Pernot P.,  2012, Icarus, 219, 254

\bibitem[\protect\citeauthoryear{{Plume}, {Jaffe}, {Evans},
  {Mart{\'{\i}}n-Pintado}  \& {G{\'o}mez-Gonz{\'a}lez}}{{Plume}
  et~al.}{1997}]{Plume97}
{Plume} R.,  {Jaffe} D.~T.,  {Evans} II N.~J.,  {Mart{\'{\i}}n-Pintado} J.,
  {G{\'o}mez-Gonz{\'a}lez} J.,  1997, \apj, \href
  {http://adsabs.harvard.edu/abs/1997ApJ...476..730P} {476, 730}

\bibitem[\protect\citeauthoryear{{Podio} et~al.,}{{Podio}
  et~al.}{2015}]{Podio15}
{Podio} L.,  et~al., 2015, \mn@doi [\aap] {10.1051/0004-6361/201525778}, \href
  {http://adsabs.harvard.edu/abs/2015A%26A...581A..85P} {581, A85}

\bibitem[\protect\citeauthoryear{{Prasad} \& {Tarafdar}}{{Prasad} \&
  {Tarafdar}}{1983}]{Prasad83}
{Prasad} S.~S.,  {Tarafdar} S.~P.,  1983, \mn@doi [\apj] {10.1086/160896},
  \href {http://adsabs.harvard.edu/abs/1983ApJ...267..603P} {267, 603}

\bibitem[\protect\citeauthoryear{Puzzarini}{Puzzarini}{2005}]{Puzzarini05}
Puzzarini C.,  2005, \mn@doi [J. Chem. Phys.]
  {doi:http://dx.doi.org/10.1063/1.1953367}, 123, 024313

\bibitem[\protect\citeauthoryear{Reiter \& Janev}{Reiter \&
  Janev}{2010}]{Reiter10}
Reiter D.,  Janev R.,  2010, Contributions to Plasma Physics, 50, 986

\bibitem[\protect\citeauthoryear{Rice \& Chabalowski}{Rice \&
  Chabalowski}{1994}]{Rice94}
Rice B.~M.,  Chabalowski C.~F.,  1994, The Journal of Physical Chemistry, 98,
  9488

\bibitem[\protect\citeauthoryear{Rice, Cartland  \& Chabalowski}{Rice
  et~al.}{1993}]{Rice93}
Rice B.~M.,  Cartland H.~E.,   Chabalowski C.~F.,  1993, Chemical physics
  letters, 211, 283

\bibitem[\protect\citeauthoryear{{Ruaud}, {Wakelam}  \& {Hersant}}{{Ruaud}
  et~al.}{2016}]{Ruaud16}
{Ruaud} M.,  {Wakelam} V.,   {Hersant} F.,  2016, \mn@doi [\mnras]
  {10.1093/mnras/stw887}, \href
  {http://adsabs.harvard.edu/abs/2016MNRAS.tmp..679R} {}

\bibitem[\protect\citeauthoryear{{Ruffle}, {Hartquist}, {Caselli}  \&
  {Williams}}{{Ruffle} et~al.}{1999}]{Ruffle99}
{Ruffle} D.~P.,  {Hartquist} T.~W.,  {Caselli} P.,   {Williams} D.~A.,  1999,
  \mn@doi [\mnras] {10.1046/j.1365-8711.1999.02562.x}, \href
  {http://adsabs.harvard.edu/abs/1999MNRAS.306..691R} {306, 691}

\bibitem[\protect\citeauthoryear{Sabbah, Biennier, Sims, Georgievskii,
  Klippenstein  \& Smith}{Sabbah et~al.}{2007}]{Sabbah07}
Sabbah H.,  Biennier L.,  Sims I.~R.,  Georgievskii Y.,  Klippenstein S.~J.,
  Smith I.~W.,  2007, Science, 317, 102

\bibitem[\protect\citeauthoryear{{Sakai} et~al.,}{{Sakai}
  et~al.}{2014}]{Sakai14}
{Sakai} N.,  et~al., 2014, \mn@doi [\nat] {10.1038/nature13000}, \href
  {http://adsabs.harvard.edu/abs/2014Natur.507...78S} {507, 78}

\bibitem[\protect\citeauthoryear{Sander et~al.,}{Sander
  et~al.}{2011}]{Sander11}
Sander S.,  et~al., 2011, JPL Publication, 10-6

\bibitem[\protect\citeauthoryear{Scott, Fairley, Freeman, McEwan  \&
  Anicich}{Scott et~al.}{1999}]{Scott99}
Scott G.~B.,  Fairley D.~A.,  Freeman C.~G.,  McEwan M.~J.,   Anicich V.~G.,
  1999, The Journal of Physical Chemistry A, 103, 1073

\bibitem[\protect\citeauthoryear{Scott, Milligan, Fairley, Freeman  \&
  McEwan}{Scott et~al.}{2000}]{Scott00}
Scott G.~B.,  Milligan D.~B.,  Fairley D.~A.,  Freeman C.~G.,   McEwan M.~J.,
  2000, The Journal of Chemical Physics, 112, 4959

\bibitem[\protect\citeauthoryear{{Shalabiea}}{{Shalabiea}}{2001}]{Shalabiea01}
{Shalabiea} O.~M.,  2001, \mn@doi [\aap] {10.1051/0004-6361:20010323}, \href
  {http://adsabs.harvard.edu/abs/2001A%26A...370.1044S} {370, 1044}

\bibitem[\protect\citeauthoryear{Shannon, Blitz, Goddard  \& Heard}{Shannon
  et~al.}{2013}]{Shannon13}
Shannon R.~J.,  Blitz M.~A.,  Goddard A.,   Heard D.~E.,  2013, Nature
  chemistry, 5, 745

\bibitem[\protect\citeauthoryear{{Smith}}{{Smith}}{1991}]{Smith91}
{Smith} R.~G.,  1991, \mn@doi [\mnras] {10.1093/mnras/249.1.172}, \href
  {http://adsabs.harvard.edu/abs/1991MNRAS.249..172S} {249, 172}

\bibitem[\protect\citeauthoryear{Smith}{Smith}{2006}]{Smith06}
Smith I.~W.,  2006, Angewandte Chemie International Edition, 45, 2842

\bibitem[\protect\citeauthoryear{{Smith}}{{Smith}}{2011}]{Smith11}
{Smith} I.~W.~M.,  2011, \mn@doi [\araa] {10.1146/annurev-astro-081710-102533},
  \href {http://adsabs.harvard.edu/abs/2011ARA%26A..49...29S} {49, 29}

\bibitem[\protect\citeauthoryear{{Sofia}, {Cardelli}  \& {Savage}}{{Sofia}
  et~al.}{1994}]{Sofia94}
{Sofia} U.~J.,  {Cardelli} J.~A.,   {Savage} B.~D.,  1994, \mn@doi [\apj]
  {10.1086/174438}, \href {http://adsabs.harvard.edu/abs/1994ApJ...430..650S}
  {430, 650}

\bibitem[\protect\citeauthoryear{{Tieftrunk}, {Pineau des Forets}, {Schilke}
  \& {Walmsley}}{{Tieftrunk} et~al.}{1994}]{Tieftrunk94}
{Tieftrunk} A.,  {Pineau des Forets} G.,  {Schilke} P.,   {Walmsley} C.~M.,
  1994, \aap, \href {http://adsabs.harvard.edu/abs/1994A%26A...289..579T} {289,
  579}

\bibitem[\protect\citeauthoryear{Tielens \& Hagen}{Tielens \&
  Hagen}{1982}]{Tielens82}
Tielens A.,  Hagen W.,  1982, Astronomy and Astrophysics, 114, 245

\bibitem[\protect\citeauthoryear{{Viti}, {Caselli}, {Hartquist}  \&
  {Williams}}{{Viti} et~al.}{2001}]{Viti01}
{Viti} S.,  {Caselli} P.,  {Hartquist} T.~W.,   {Williams} D.~A.,  2001,
  \mn@doi [\aap] {10.1051/0004-6361:20010300}, \href
  {http://adsabs.harvard.edu/abs/2001A%26A...370.1017V} {370, 1017}

\bibitem[\protect\citeauthoryear{Wagner \& Bowman}{Wagner \&
  Bowman}{1987}]{Wagner87}
Wagner A.~F.,  Bowman J.~M.,  1987, Journal of Physical Chemistry, 91, 5314

\bibitem[\protect\citeauthoryear{{Wakelam} \& {Herbst}}{{Wakelam} \&
  {Herbst}}{2008}]{Wakelam08}
{Wakelam} V.,  {Herbst} E.,  2008, \mn@doi [\apj] {10.1086/587734}, \href
  {http://adsabs.harvard.edu/abs/2008ApJ...680..371W} {680, 371}

\bibitem[\protect\citeauthoryear{{Wakelam}, {Caselli}, {Ceccarelli}, {Herbst}
  \& {Castets}}{{Wakelam} et~al.}{2004}]{Wakelam04}
{Wakelam} V.,  {Caselli} P.,  {Ceccarelli} C.,  {Herbst} E.,   {Castets} A.,
  2004, \mn@doi [\aap] {10.1051/0004-6361:20047186}, \href
  {http://adsabs.harvard.edu/abs/2004A%26A...422..159W} {422, 159}

\bibitem[\protect\citeauthoryear{{Wakelam}, {Herbst}  \& {Selsis}}{{Wakelam}
  et~al.}{2006}]{Wakelam06}
{Wakelam} V.,  {Herbst} E.,   {Selsis} F.,  2006, \mn@doi [\aap]
  {10.1051/0004-6361:20054682}, \href
  {http://adsabs.harvard.edu/abs/2006A%26A...451..551W} {451, 551}

\bibitem[\protect\citeauthoryear{Wakelam et~al.,}{Wakelam
  et~al.}{2010}]{Wakelam10}
Wakelam V.,  et~al., 2010, Space science reviews, 156, 13

\bibitem[\protect\citeauthoryear{{Wakelam} et~al.,}{{Wakelam}
  et~al.}{2012}]{Wakelam12}
{Wakelam} V.,  et~al., 2012, \mn@doi [\apjs] {10.1088/0067-0049/199/1/21},
  \href {http://adsabs.harvard.edu/abs/2012ApJS..199...21W} {199, 21}

\bibitem[\protect\citeauthoryear{{Wakelam} et~al.,}{{Wakelam}
  et~al.}{2015a}]{Wakelam15}
{Wakelam} V.,  et~al., 2015a, \mn@doi [\apjs] {10.1088/0067-0049/217/2/20},
  \href {http://adsabs.harvard.edu/abs/2015ApJS..217...20W} {217, 20}

\bibitem[\protect\citeauthoryear{{Wakelam}, {Loison}, {Hickson}  \&
  {Ruaud}}{{Wakelam} et~al.}{2015b}]{Wakelam15b}
{Wakelam} V.,  {Loison} J.-C.,  {Hickson} K.~M.,   {Ruaud} M.,  2015b, \mn@doi
  [\mnras] {10.1093/mnrasl/slv097}, \href
  {http://adsabs.harvard.edu/abs/2015MNRAS.453L..48W} {453, L48}

\bibitem[\protect\citeauthoryear{Wang, Eyre  \& Dorfman}{Wang
  et~al.}{1973}]{Wang73}
Wang H.,  Eyre J.,   Dorfman L.~M.,  1973, The Journal of Chemical Physics, 59,
  5199

\bibitem[\protect\citeauthoryear{Wierzejewska \& Moc}{Wierzejewska \&
  Moc}{2003}]{Wierzejewska03}
Wierzejewska M.,  Moc J.,  2003, The Journal of Physical Chemistry A, 107,
  11209

\bibitem[\protect\citeauthoryear{Woods, Occhiogrosso, Viti,
  Ka{\v{n}}uchov{\'a}, Palumbo  \& Price}{Woods et~al.}{2015}]{Woods15}
Woods P.~M.,  Occhiogrosso A.,  Viti S.,  Ka{\v{n}}uchov{\'a} Z.,  Palumbo
  M.~E.,   Price S.~D.,  2015, Monthly Notices of the Royal Astronomical
  Society, 450, 1256

\bibitem[\protect\citeauthoryear{Woon \& Herbst}{Woon \& Herbst}{2009}]{Woon09}
Woon D.~E.,  Herbst E.,  2009, The Astrophysical Journal Supplement Series,
  185, 273

\bibitem[\protect\citeauthoryear{Yoshimura, Koshi, Matsui, Kamiya  \&
  Umeyama}{Yoshimura et~al.}{1992}]{Yoshimura92}
Yoshimura M.,  Koshi M.,  Matsui H.,  Kamiya K.,   Umeyama H.,  1992, Chemical
  physics letters, 189, 199

\bibitem[\protect\citeauthoryear{Zhao \& Truhlar}{Zhao \&
  Truhlar}{2008}]{Zhao08}
Zhao Y.,  Truhlar D.~G.,  2008, Journal of Chemical Theory and Computation, 4,
  1849

\makeatother
\end{thebibliography}




\bsp	
\label{lastpage}
\end{document}